\documentclass[a4paper,11pt]{article}
\usepackage{jcappub} 
\usepackage{lineno}
\usepackage{cleveref,subcaption}

\title{\boldmath \texttt{DSWIM:}Efficient and Stable Deterministic Computation of Warm Inflation Perturbations}

\author{Umang Kumar}
\affiliation{Department of Physics, Ashoka University, Rajiv Gandhi Education City, Rai, Sonipat:
131029, Haryana, India}

\emailAdd{umang.kumar\_phd21@ashoka.edu.in}

\abstract{
Warm inflation perturbations are sourced by both thermal and quantum fluctuations and are commonly computed through stochastic realizations of the perturbation equations, as implemented in the publicly available code \texttt{SWIM}. Deterministic formulations based on correlation matrix evolution provide a computationally efficient alternative, but can become numerically ill-conditioned when the perturbation variables evolve over widely different scales. In this work, we extend \texttt{SWIM} by introducing a deterministic module, \texttt{DSWIM}, based on correlation matrix evolution. We introduce a physically motivated scaling matrix transformation derived from the effective Hubble scaling of the perturbation variables. The transformed system preserves the primordial curvature power spectrum exactly while substantially improving the numerical conditioning of the deterministic evolution equations. Using representative warm inflation models, we show that the scaled framework suppresses numerical artifacts, improves the robustness of the deterministic evolution, and yields substantial computational speedups while preserving accuracy. We further show that correlated thermal noise contributions arise naturally through the diffusion matrix structure, resolving previously observed discrepancies between stochastic and deterministic implementations. Our results establish \texttt{DSWIM} as a numerically robust and computationally efficient framework for computing warm inflation scalar perturbations.
}

\begin{document}
\maketitle
\flushbottom

\section{Introduction}
\label{sec:intro}

Inflation~\cite{Guth:1980zm} provides a compelling mechanism for resolving several shortcomings of the standard Big Bang cosmology while simultaneously generating the primordial density perturbations that seed the observed large-scale structure of the Universe. In the standard cold inflation paradigm, the accelerated expansion is driven by the potential energy of a scalar inflaton field and is followed by a separate reheating phase during which the inflaton transfers its energy into radiation. Warm inflation (WI) \cite{Berera:1995ie} provides an alternative framework in which dissipative interactions continuously convert inflaton energy into radiation during inflation itself, allowing the Universe to smoothly transition into the radiation-dominated era without requiring a separate reheating phase.

In WI, the inflaton evolves in the presence of a thermal radiation bath with temperature satisfying $T>H$, where $H$ is the Hubble parameter. Dissipative effects introduce an additional friction term in the inflaton dynamics, permitting slow-roll inflation even for comparatively steep potentials \cite{Das:2020xmh,Kumar:2024hju,Das:2025teu}. As a consequence, WI can accommodate a broad class of inflationary potentials that are otherwise difficult to realize in conventional cold inflation scenarios~\cite{Bastero-Gil:2019gao,Bartrum:2013fia,Berghaus:2025dqi,ORamos:2025uqs}. Warm inflation has also been shown to alleviate tensions with the swampland conjectures \cite{Das:2018rpg,Motaharfar:2018zyb,Das:2019hto}, naturally suppress the tensor-to-scalar ratio, and provide viable realizations within particle physics motivated constructions such as Warm Little Inflation \cite{Bastero-Gil:2016qru}, Minimal Warm Inflation \cite{Berghaus:2019whh}, and effective field theory based realizations \cite{Bastero-Gil:2019gao}. Recent analyses have further demonstrated that several WI models remain compatible with current cosmological observations \cite{Das:2025teu,Berera:2025vsu,Chakraborty:2025yms,Chakraborty:2026eep,Chakraborty:2025jof} by PLANCK \cite{Planck:2018vyg,Planck:2018jri} and ACT \cite{AtacamaCosmologyTelescope:2025blo,AtacamaCosmologyTelescope:2025nti}. For recent reviews on WI, see \cite{Kamali:2023lzq,Berera:2023liv}.

A distinctive feature of WI is the presence of dissipative and stochastic effects in the perturbation dynamics. In addition to the standard quantum fluctuations of the inflaton, thermal fluctuations sourced by the radiation bath can significantly contribute to the primordial scalar power spectrum~\cite{Berera:1995wh,Taylor:2000ze}. Consequently, the perturbation evolution is governed by a coupled stochastic system~\cite{Hall:2003zp,Ramos:2013nsa,Bastero-Gil:2011rva} involving inflaton, radiation and metric perturbations. In particle physics realizations of WI, the dissipation coefficient $\Upsilon$ generally depends on temperature and the inflaton field itself, which couples inflaton and radiation perturbations and renders analytical treatments highly non-trivial. Although approximate analytical expressions for the scalar power spectrum can be derived under the assumptions of slow-roll evolution and a temperature independent dissipation coefficient \cite{Graham_2009,Ramos:2013nsa}, the latter assumption is not generally satisfied in realistic WI models, where the dissipation coefficient typically exhibits an explicit temperature dependence \cite{Berghaus:2019whh,Bastero-Gil:2016qru,Bastero-Gil:2019gao,Berera:2025vsu}. The effects of this temperature dependence are conventionally encoded through a correction factor $G(Q)$ multiplying the analytical spectrum where $Q$ denotes the dissipation ratio. Physically, $Q$ measures the relative strength of dissipative friction compared to Hubble friction in the inflaton dynamics and therefore characterizes the strength of the coupling between the inflaton and the radiation bath. This distinction naturally divides WI into the weak dissipative regime ($Q\ll1$), where Hubble friction dominates, the strong dissipative regime ($Q\gg1$), where dissipative effects dominate the dynamics~\cite{Berera:2008ar}, and an intermediate (moderate dissipative) regime ($0.01\lesssim Q\lesssim10$) where both contributions can play an important role. 

The functional form of $G(Q)$ generally depends on the temperature scaling of the dissipation coefficient and weakly on the form of the inflaton potential~\cite{Montefalcone:2023pvh}. In practice, $G(Q)$ is obtained from the numerical evolution of the full perturbation system and is typically represented through fitting functions or interpolated tabulated data. Consequently, even semi-analytical calculations of the WI power spectrum ultimately rely on numerical evolution of the perturbation system. The standard numerical approach to WI perturbations is based on stochastic realizations of the coupled perturbation system, followed by ensemble averaging to determine the scalar power spectrum. Recently, this framework was implemented in the publicly available code \texttt{SWIM}~\cite{Kumar:2026mvz}, which numerically evolves the full stochastic perturbation equations without relying on the approximations employed in semi-analytical treatments.\footnote{\texttt{WarmSPy}~\cite{Montefalcone:2023pvh} previously implemented a stochastic framework for computing the correction factor $G(Q)$. However, it is restricted to specific warm inflation models, assumes a constant dissipation ratio, and relies on fitted analytical representations of $G(Q)$ that may not accurately capture its full behaviour in more general scenarios. For a detailed comparison between \texttt{WarmSPy} and \texttt{SWIM}, see section~3 of ref.~\cite{Kumar:2026mvz}.} While this stochastic formulation is general and robust, obtaining statistically converged spectra requires averaging over a large number of realizations~\cite{Ballesteros:2023dno,Kumar:2026mvz}, making the computation computationally expensive, particularly in applications involving repeated likelihood evaluations and parameter inference.

An alternative deterministic formulation based on the Fokker--Planck approach was developed in \cite{Ballesteros:2022hjk,Ballesteros:2023dno}, where its equivalence with the stochastic treatment was also demonstrated. Subsequently, this formalism was implemented in the publicly available code \texttt{WI2easy} \cite{Rodrigues:2025neh}. Instead of evolving stochastic realizations directly, this formalism evolves the correlation matrix of the perturbation variables through a coupled deterministic system of ordinary differential equations. Since only a single deterministic evolution is required for each comoving mode, this approach can substantially reduce the computational cost relative to the stochastic formulation.

However, in this work we show that the deterministic formulation can become severely numerically ill-conditioned for certain WI models. Since the perturbation variables exhibit different scaling behaviour during inflation, the elements of the correlation matrix can span many orders of magnitude during the evolution, causing some components to approach machine precision while others remain comparatively large. This leads to floating-point inaccuracies, artificial distortions in the correction factor $G(Q)$, and eventual breakdown of the deterministic evolution for sufficiently extreme models. We demonstrate that these numerical instabilities arise from the conditioning of the correlation matrix evolution rather than from the deterministic formalism itself.

To address this issue, we develop a scaling matrix framework for the deterministic evolution equations and implement it as a separate deterministic module within the publicly available code \texttt{SWIM}. We refer to this implementation as \texttt{DSWIM} (Deterministic (version of) \texttt{SWIM}) throughout this work.\footnote{\texttt{SWIM}, including the deterministic module \texttt{DSWIM} developed in this work, is publicly available at \url{https://github.com/umg-kmr/SWIM}.}  While \texttt{SWIM} was originally developed as a stochastic WI perturbation solver, the present work extends its capabilities to include stable deterministic evolution of the perturbation correlation matrix. The scaling transformation is physically motivated by the effective scaling behaviour of the perturbation variables and substantially improves the numerical conditioning of the deterministic system while preserving the primordial scalar power spectrum exactly. Using representative WI models, we demonstrate that the scaled deterministic implementation suppresses numerical artifacts, significantly extends the stable evolution range, and remains in excellent agreement with the stochastic formulation. Throughout this work, we refer to the original deterministic formulation~\cite{Ballesteros:2022hjk,Ballesteros:2023dno,Rodrigues:2025neh} without the scaling transformation as the \emph{unscaled} deterministic approach and to the present implementation incorporating the scaling framework as the \emph{scaled} deterministic approach.

We additionally discuss the role of correlated thermal noise contributions in the deterministic formalism and show that incorporating these correlations resolves the systematic discrepancy previously observed between stochastic \texttt{SWIM} and \texttt{WI2easy} in the moderate dissipative regime.

This paper is organized as follows. In \cref{sec:theoretical_bg} we briefly review the stochastic perturbation equations in WI and the deterministic Fokker--Planck formalism. In \cref{sec:S_Formalism} we discuss the numerical challenges associated with the deterministic evolution and introduce the scaling matrix framework. The numerical results and comparisons with stochastic \texttt{SWIM} and \texttt{WI2easy} are presented in \cref{sec:Results}. Finally, we summarize our conclusions in \cref{sec:Discussion}.

\section{Theoretical background}
\label{sec:theoretical_bg}
In this section, we briefly review the stochastic perturbation equations governing scalar perturbations in WI and the corresponding deterministic formulation based on the Fokker-Planck approach. We first present the coupled stochastic evolution equations for the inflaton fluctuations, perturbed radiation energy density, momentum perturbation of the radiation bath, and the scalar metric perturbations. We then discuss how the stochastic perturbation system can be reformulated as a deterministic evolution equation following~\cite{Ballesteros:2022hjk,Ballesteros:2023dno,Rodrigues:2025neh} of the correlation matrix of the perturbation variables. This deterministic formulation forms the basis for the scaling framework developed later in this work.

\subsection{Background Equations in WI}
\label{sec:WI_bg}
In WI, dissipative interactions continuously transfer energy from the inflaton to a sub--dominant radiation bath during inflation. The background evolution is therefore governed by a coupled inflaton--radiation system in which dissipative effects continuously source the radiation bath during inflation and smoothly transitions to a radiation dominated (RD) phase after inflation ends. The equation of motion for the homogeneous inflaton field is given by
\begin{equation}
\ddot{\phi} + \left(3H + \Upsilon\right)\dot{\phi} + V_{,\phi} = 0 \, ,
\label{eq:KG}
\end{equation}
where an overdot denotes differentiation with respect to cosmic time, $V_{,\phi} \equiv \partial V/\partial\phi$, and $\Upsilon \equiv \Upsilon(\phi,T)$ is the dissipation coefficient whose functional form depends on the underlying microphysics of the WI model under consideration. The dissipative transfer of energy from the inflaton to radiation leads to the radiation continuity equation
\begin{equation}
\dot{\rho}_r + 4 H \rho_r = \Upsilon \dot{\phi}^2 \, ,
\label{eq:radn_contn}
\end{equation}
where the radiation energy density is given by
\begin{equation}
\rho_r = \frac{\pi^2}{30} g_* T^4 \, ,
\end{equation}
with $g_*$ denoting the effective number of relativistic degrees of freedom of the thermal radiation bath and $T$ the corresponding equilibrium temperature.

The system of background evolution equations is completed by the Friedmann equation,
\begin{equation}\label{eq:Friedmann}
3 M_{\rm Pl}^2 H^2 = \frac{\dot{\phi}^2}{2} + V(\phi) + \rho_r \, ,
\end{equation}
where $M_{\rm Pl} \equiv (8\pi G)^{-1/2}$ is the reduced Planck mass. Together,~\cref{eq:KG,eq:radn_contn} and the Friedmann equation form the coupled background system governing the evolution of the inflaton field, radiation bath and cosmic expansion in WI. Eq.~\eqref{eq:KG} shows that dissipation acts as an additional friction term in the inflaton evolution. The relative strength of dissipative friction compared to Hubble friction is quantified by the dimensionless dissipation ratio

\begin{equation}
Q \equiv \frac{\Upsilon}{3H} \, .
\label{eq:Q}
\end{equation}

Depending on the magnitude of $Q$, WI scenarios can broadly be classified into two regimes:

\begin{itemize}
    \item[(i)] the weak dissipative regime (WDR), characterized by $Q \ll 1$, and
    \item[(ii)] the strong dissipative regime (SDR), characterized by $Q \gg 1$.
\end{itemize}

In the WDR, the Hubble friction term dominates the inflaton dynamics, and the background evolution can closely resemble that of CI. However, the presence of a thermal bath with temperature satisfying $T > H$ remains the defining condition for the dynamics to be warm. The functional form of the dissipative coefficient $\Upsilon(\phi,T)$ depends on the underlying microphysics of the WI model under consideration and can, in general, be parameterized as
\begin{equation}
\Upsilon(\phi,T) = C_{\Upsilon} T^p \phi^c \, ,
\label{eq:gen_Ups}
\end{equation}
where $C_{\Upsilon}$ is a model-dependent constant whose magnitude and dimensions are determined by the microscopic realization of WI. The exponents $p$ and $c$ are integers, with the stability analysis of WI imposing the constraint $-4 < p < 4$ \cite{Moss:2008yb,delCampo:2010by,Bastero-Gil:2012vuu,Das:2023rat}.

The parametrization in~\cref{eq:gen_Ups} captures several microphysical realizations of WI studied in the literature, including i) the Warm Little Inflaton model \cite{Bastero-Gil:2016qru}, which yields $\Upsilon \propto T$, ii) Minimal Warm Inflation \cite{Berghaus:2019whh}, where $\Upsilon \propto T^3$, and iii) the two-stage decay mechanism \cite{Moss:2006gt,Bastero-Gil:2010dgy,Bastero-Gil:2012akf}, which gives $\Upsilon \propto T^3/\phi^2$. However, not all microphysical realizations of WI can be conveniently expressed in the simple parametrized form of~\cref{eq:gen_Ups}. For example, in the effective field theory realization known as the Warm Little Inflaton Scalar (WLIS) model \cite{Bastero-Gil:2019gao,Berera:2025vsu}, the dissipative coefficient takes the form
\begin{equation}
\Upsilon \simeq Cg^4 \frac{M^2T^2}{m^3} \left[ 1 + \frac{1}{\sqrt{2\pi}} \left( \frac{m}{T} \right)^{3/2} \right] e^{-m/T} \, ,
\label{eq:EFT}
\end{equation}
where
\begin{equation}
m^2 = m_0^2 + \alpha^2 T^2 \, ,
\end{equation}
with $m_0$ denoting the vacuum mass of the light scalar fields and $\alpha$ representing the corresponding coupling constant. Similarly, WI scenarios involving Standard Model interactions can lead to dissipative coefficients of the form \cite{Berghaus:2025dqi,ORamos:2025uqs}
\begin{equation}
\Upsilon = \frac{\kappa T^3 (a_g N_c)^5}{2f_a^2} \left( \frac{1}{1+\kappa a_g^5 N_c^5 (2N_f/N_c)(T/H)} \right) \, ,
\end{equation}
where $a_g$ is the gluon gauge coupling, $\kappa$ is a dimensionless constant determined by $N_c$, $N_f$, and $f_a$ the axion decay constant. 

The background quantities determined by the system above enter directly into the stochastic perturbation equations governing scalar fluctuations in WI, which we review in the following subsection.

\subsection{Scalar Perturbations in WI}
\label{sec:pert}
In this subsection, we briefly review the stochastic scalar perturbation equations used in \texttt{SWIM} \cite{Kumar:2026mvz}. We work in the Newtonian gauge, for which the perturbed FLRW metric is given by
\begin{equation}
    \text{d}s^2 = -(1+2\psi)\text{d}t^2 + a^2(1-2\psi)\delta_{ij}\text{d}x^i\text{d}x^j \, ,
\end{equation}
where $\psi$ denotes the scalar metric perturbation and $a$ is the scale factor. Using the number of e-folds $N$ as the temporal variable, with ${}'\equiv \mathrm{d}/\mathrm{d}N$, the coupled evolution equations for the inflaton, radiation, momentum and metric perturbations can be written as~\cite{Kumar:2026mvz,Rodrigues:2025neh,Montefalcone:2023pvh,Ballesteros:2023dno,Ballesteros:2022hjk,Kamali:2023lzq}
\begin{align}
    &\delta\phi'' = -\left( 3 + \frac{\Upsilon}{H} + \frac{H'}{H} \right)\delta\phi' -\left(  \frac{k^2}{a^2 H^2}  + \frac{V_{,\phi\phi}}{H^2} + \frac{\Upsilon_{,\phi}\phi'}{H} \right)\delta \phi -\frac{\Upsilon_{,T}T\phi'\delta\rho_r}{4H\rho_r} + 4\psi'\phi'
    \nonumber\\[4pt]
    \label{eq:pert_KG} 
    &\phantom{\delta\phi'' =}
    -\left( \frac{\Upsilon\phi'}{H} + \frac{2V_{,\phi}}{H^2} \right)\psi + \sqrt{ \frac{2\Upsilon T}{a^3H^3}}\xi_T + ( 9H + 4\pi\Upsilon )^{1/4} \sqrt{\frac{ 1 + 2n }{\pi a^3H^{3/2}}}\xi_q \, , \\ 
    &\delta \rho_r' = - \left(4 -  \frac{\Upsilon_{,T}H\phi'^2T}{4\rho_r} \right)\delta\rho_r + \frac{k^2}{a^2H}\delta q_r + 2\Upsilon H\phi'\delta \phi' + \Upsilon_{,\phi}H\phi'^2\delta \phi + 4 \rho_r\psi' 
    \label{eq:pert_rad}
    \nonumber\\[4pt]
    &\phantom{\delta\rho_r' =}
    -\Upsilon H\phi'^2\psi - \sqrt{ \frac{2\Upsilon HT}{a^3}}\phi'\xi_T\, , \\ 
    &\delta q_r' = -3\delta q_r -\Upsilon \phi' \delta \phi -\frac{\delta \rho_r}{3H} - \frac{4\rho_r}{3H}\psi \, , \label{eq:pert_mtm} \\ 
    &\psi' = -\psi - \frac{1}{2 M_{\rm Pl}^2} \left( \frac{\delta q_r}{H} -\phi'\delta \phi  \right)  \, , \label{eq:pert_m}
\end{align}
where $\Upsilon_{,\phi}\equiv \partial\Upsilon/\partial\phi$, and $\Upsilon_{,T}\equiv \partial\Upsilon/\partial T$. Here, $\delta q_r$ denotes the momentum perturbation of the radiation bath and $n$ represents the statistical distribution of the inflaton fluctuations. In the case where the inflaton thermalizes with the radiation bath, $n$ is typically taken to follow the Bose--Einstein distribution. The stochastic sources $\xi_T$ and $\xi_q$ denote the thermal and quantum noise terms respectively,\footnote{The normalization of the quantum noise term $\xi_q$ is chosen such that the resulting solution reproduces the correct primordial power spectrum. Consequently, when this quantum stochastic source is retained, the inflaton perturbations may be initialized with vanishing initial conditions~\cite{Rodrigues:2025neh} as done in this work as well as~\cite{Kumar:2026mvz}. Alternative prescriptions for incorporating the quantum contribution have also been adopted in the literature~\cite{Ballesteros:2022hjk,Ballesteros:2023dno,Rodrigues:2025neh}. As discussed in~\cite{Ballesteros:2023dno}, these different prescriptions can lead to quantitative differences in the moderate dissipative regime. A detailed comparison of the various approaches is presented in appendix~\ref{appendix:quantum-noise}.} satisfying

\begin{eqnarray}
    &&\langle\xi_T (k, N)\xi_T (k', N')\rangle = \delta(N - N')(2\pi)^3 \delta(k - k') \, , \\
    &&\langle\xi_q (k, N)\xi_q (k', N')\rangle = \delta(N - N')(2\pi)^3 \delta(k - k') \, .
\end{eqnarray}

The~\cref{eq:pert_KG,eq:pert_rad,eq:pert_mtm,eq:pert_m}
form the complete set of scalar perturbation equations in the presence of both thermal and quantum stochastic fluctuations. The power spectrum of the comoving curvature perturbation, $\mathcal{R}$, is defined as
\begin{equation}
P_{\mathcal{R}} = \frac{k^3}{2\pi^2} \left. \langle |\mathcal{R}|^2 \rangle \right|_{k\ll aH} \, , \label{eq:power_spectr}
\end{equation}
where $\langle \cdots \rangle$ denotes the ensemble average over multiple realizations of the stochastic perturbations. For a multicomponent cosmological system such as WI, the comoving curvature perturbation is given by

\begin{equation}
\mathcal{R} = \frac{H}{\bar{\rho}+\bar{p}} \left( \delta q_r - H\phi'\delta\phi\right) - \psi \, ,\label{eq:comoving_R}
\end{equation}
where $\bar{\rho}$ and $\bar{p}$ denote the total background energy density and pressure of the inflaton-radiation system respectively.

Assuming that the dissipative coefficient $\Upsilon$ has no explicit temperature dependence, corresponding to $p=0$, and working within the slow-roll approximation, one obtains the following analytical approximation for the primordial scalar power spectrum in WI~\cite{Ramos:2013nsa}:
\begin{equation}
P_{\mathcal{R},\text{analytical}} = \left(\frac{H}{2\pi\phi'}\right)^2\left[1+2n+\frac{T}{H}\frac{2\sqrt{3}\pi Q}{\sqrt{3+4\pi Q}}\right] \, , \label{eq:an_Ps}
\end{equation}
where $n$ denotes the statistical distribution of the inflaton fluctuations. Since the expression in~\cref{eq:an_Ps} is derived under simplifying assumptions and neglects any explicit temperature dependence of $\Upsilon$, a multiplicative correction factor $G(Q)$ must generally be determined numerically by solving the full perturbation system given in~\cref{eq:pert_KG,eq:pert_rad,eq:pert_mtm,eq:pert_m}. The full numerical power spectrum can then be written as
\begin{equation}
P_{\mathcal{R},\text{numerical}} = P_{\mathcal{R},\text{analytical}} \, G(Q) \, .
\end{equation}
Although fitting functions for $G(Q)$ have been obtained for a number of representative WI models~\cite{Kamali:2023lzq}, no general analytical expression is known. Consequently, $G(Q)$ must be determined by numerical evolution of the perturbation equations. So far, three publicly available codes have been developed for this purpose, namely \texttt{WarmSPy}~\cite{Montefalcone:2023pvh}, \texttt{WI2easy}~\cite{Rodrigues:2025neh}, and \texttt{SWIM}~\cite{Kumar:2026mvz}. These codes provide numerical determinations of $G(Q)$ either through fitted analytical expressions (\texttt{WarmSPy}) or through tabulated numerical data that can be interpolated to obtain a smooth representation of the correction factor (\texttt{WI2easy} and \texttt{SWIM}). It is generally understood in the WI literature that the correction factor $G(Q)$ depends primarily on the temperature scaling of the dissipative coefficient $\Upsilon$, with a comparatively weaker dependence on the inflaton potential. However, recent work in \cite{Kumar:2026mvz} has shown that $G(Q)$ can additionally exhibit a non-trivial dependence on other model parameters, such as the effective number of relativistic degrees of freedom $g_*$ and the overall normalization of the inflaton potential $V_0$. In such cases, a single universal fitting function for $G(Q)$ may be insufficient, requiring repeated numerical evaluations across the relevant parameter space. Moreover, as discussed in \cite{Kumar:2026mvz}, the correction-factor approach can be bypassed altogether by directly computing the full numerical power spectrum during parameter inference. Both approaches therefore rely on repeated solutions of the perturbation equations, making computational efficiency a key consideration. This provides a strong motivation for developing efficient deterministic formulations of WI perturbations, which we discuss in the following subsection.

\subsection{Matrix Formalism and the Fokker-Planck Approach} \label{sec:FP_matrixForm}

In the previous subsection, we reviewed the stochastic formulation of scalar perturbations in WI. Since the primordial power spectrum is obtained through ensemble averaging over many stochastic realizations, direct stochastic evolution can become computationally expensive. An alternative deterministic formulation based on the Fokker--Planck approach was developed in \cite{Ballesteros:2022hjk,Ballesteros:2023dno} and implemented in the publicly available code \texttt{WI2easy} \cite{Rodrigues:2025neh}. In this framework, the stochastic perturbation system is recast into a deterministic evolution equation for the statistical moments of the perturbation variables. The coupled stochastic perturbation system is first written in the form of a matrix Langevin equation,
\begin{equation}
\mathbf{\Phi}' = \mathbf{A}\mathbf{\Phi} + \mathbf{B}_{T}\xi_T + \mathbf{B}_{q}\xi_q \, , \label{eq:matrix_langevin}
\end{equation}
where $\mathbf{\Phi}$ denotes the perturbation state vector, 
\begin{equation}
\mathbf{\Phi} =
\begin{pmatrix}
\psi \\
\delta q_r \\
\delta \phi \\
\delta \rho_r \\
\delta \phi'
\end{pmatrix} \, ,
\label{eq:state_vector}
\end{equation}
$\mathbf{A}$ is a real $5\times5$ matrix, and $\mathbf{B}_{T}$ and $\mathbf{B}_{q}$ are real column vectors corresponding to the thermal and quantum stochastic noise sources respectively. The explicit forms of the matrix $\mathbf{A}$ and the noise vectors $\mathbf{B}_{T}$ and $\mathbf{B}_{q}$ can be obtained by comparing \cref{eq:matrix_langevin} with the perturbation system given in \cref{eq:pert_KG,eq:pert_rad,eq:pert_mtm,eq:pert_m}, together with the definition of the state vector in \cref{eq:state_vector}. The explicit matrix elements are presented in the appendix~\ref{appendix:matrices}.

Following \cite{Ballesteros:2022hjk}, the primordial scalar power spectrum can be written in matrix form as
\begin{equation}
P_{\mathcal{R}}  = \frac{k^3}{2\pi^2} \mathbf{C}^{T} \mathbf{J} \mathbf{C} \, , \label{eq:matrx_PS}
\end{equation}
where $\mathbf{J}$
\begin{equation}\label{eq:J_correlation}
\mathbf{J} \equiv \left\langle \mathbf{\Phi}\mathbf{\Phi}^{\dagger} \right\rangle\, ,
\end{equation}
is the correlation matrix of the perturbation variables encoding the stochastic dynamics in the evolution of the quadratic statistical moments. Explicitly, the correlation matrix is defined as
\begin{equation}
\left\langle \mathbf{\Phi}\mathbf{\Phi}^{\dagger} \right\rangle (N) = \int \prod_i \text{d}\mathbf{\Phi}_i \int \prod_j \text{d}\mathbf{\Phi}^{*}_j \, P(\mathbf{\Phi},\mathbf{\Phi}^{*},N) \, \mathbf{\Phi}\mathbf{\Phi}^{\dagger} \, . \label{eq:statistical_moment}
\end{equation}
The comoving curvature perturbation can similarly be written as
\begin{equation}
\mathcal{R} = \mathbf{C}^{T}\mathbf{\Phi} \, . \label{eq:matrx_R}
\end{equation}
 The explicit form of $\mathbf{C}$ can be obtained directly by comparing \cref{eq:matrx_R} with the definition of $\mathcal{R}$ given in \cref{eq:comoving_R}. The components of $\mathbf{C}$ are presented explicitly in the appendix~\ref{eq:C-matrix_explicit}. The evolution of the probability density $P(\mathbf{\Phi},\mathbf{\Phi}^{*},N)$ is governed by the Fokker-Planck equation
\begin{equation}
\frac{\partial P}{\partial N} = \sum_{kl} \left[ \mathbf{A}_{kl} \frac{\partial}{\partial \mathbf{\Phi}_k} \left( \mathbf{\Phi}_l P \right) + \mathbf{A}_{kl} \frac{\partial}{\partial \mathbf{\Phi}^{*}_k} \left( \mathbf{\Phi}^{*}_l P \right) + \mathbf{D}_{kl} \frac{\partial^2 P}{ \partial \mathbf{\Phi}_k \partial \mathbf{\Phi}^{*}_l } \right] \, , \label{eq:FP}
\end{equation}
where the diffusion matrix $\mathbf{D}$ encodes the statistical correlations of the stochastic noise contributions (thermal and quantum) and is given by
\begin{equation}\label{eq:diffusion_matrix}
\mathbf{D} = \mathbf{B}_T\mathbf{B}_T^{T} + \mathbf{B}_q\mathbf{B}_q^{T} \, ,
\end{equation}
following from the assumption that the thermal and quantum noise sources are statistically uncorrelated. Correlations between thermal noise contributions entering different perturbation equations are encoded through the structure of $\mathbf{D}$. The matrix $\mathbf{D}$ is real-valued, and its explicit components are presented in the appendix~\ref{eq:D-matrix_explicit}.

The stochastic evolution problem can then be replaced by a deterministic evolution equation for the correlation matrix $\mathbf{J}$. Differentiating~\cref{eq:statistical_moment} with respect to the number of e-folds $N$ and using the Fokker--Planck~\cref{eq:FP} yields\footnote{Equation~\eqref{eq:FP_ode} differs in sign from the corresponding expression in refs.~\cite{Ballesteros:2022hjk,Ballesteros:2023dno,Rodrigues:2025neh} due to a different convention for the drift matrix. In the present work the perturbation system is written as $\Phi'=\mathbf A\Phi+\mathbf{B}\xi$, whereas refs.~\cite{Ballesteros:2022hjk,Ballesteros:2023dno,Rodrigues:2025neh} adopt the convention $\Phi'+\mathbf A\Phi=\mathbf{B}\xi$. }
\begin{equation}
\mathbf{J}' = \mathbf{A}\mathbf{J} + \mathbf{J}\mathbf{A}^{T} + \mathbf{D}\, .
\label{eq:FP_ode}
\end{equation}
The deterministic matrix~\cref{eq:FP_ode} can then be solved directly to obtain the evolution of the correlation matrix $\mathbf{J}$ and consequently the numerical WI power spectrum. This approach bypasses the need to solve the full stochastic system repeatedly over multiple realizations, leading to a substantial reduction in computational cost.

\section{Numerical Stability of the Deterministic Formalism}
\label{sec:S_Formalism}

Although the deterministic formulation of WI perturbations provides a computationally efficient alternative to direct stochastic realizations, direct evolution of the correlation matrix can suffer from significant numerical instabilities for certain WI models. In this section, we discuss the origin of these numerical difficulties and introduce a scaling matrix framework that stabilizes the deterministic evolution equations while preserving the primordial curvature power spectrum exactly.

\subsection{Challenges with the Deterministic Approach}

The deterministic formulation discussed in the previous section is computationally more efficient than the direct stochastic realization approach implemented in \texttt{SWIM}~\cite{Kumar:2026mvz} as well as \texttt{WarmSPy}~\cite{Montefalcone:2023pvh}, while remaining formally equivalent. However, evolving the deterministic system introduces several important numerical challenges.

For the five-component perturbation state vector considered here, the deterministic evolution~\cref{eq:FP_ode} corresponds to a coupled system of $25$ ordinary differential equations governing the evolution of the elements of the correlation matrix $\mathbf{J}$. A more significant challenge originates from the large hierarchy of scales among the perturbation variables. During the evolution, different components of the perturbation vector can differ by many orders of magnitude. Since the deterministic system evolves quadratic statistical moments rather than the perturbation variables themselves, the elements of $\mathbf{J}$ can become substantially smaller in magnitude during the evolution, further amplifying the hierarchy of scales. As a result, several components of $\mathbf{J}$ can dynamically approach machine precision, leading to accumulation of floating-point round-off errors and eventual degradation of numerical accuracy.

\begin{figure}[htbp]
\centering

\begin{subfigure}[t]{0.49\linewidth}
    \centering
    \includegraphics[width=\linewidth]{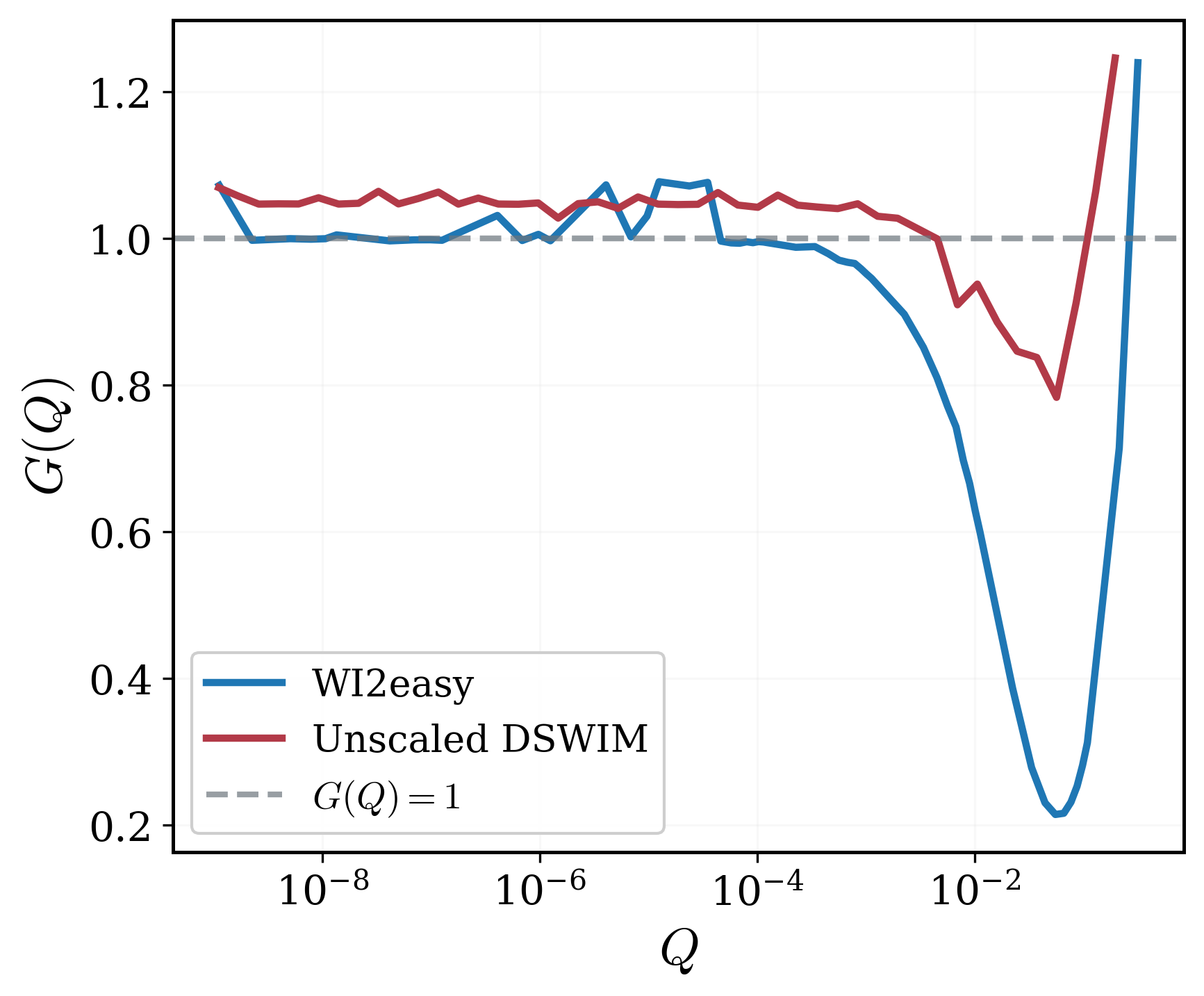}
\end{subfigure}
\hfill
\begin{subfigure}[t]{0.49\linewidth}
    \centering
    \includegraphics[width=\linewidth]{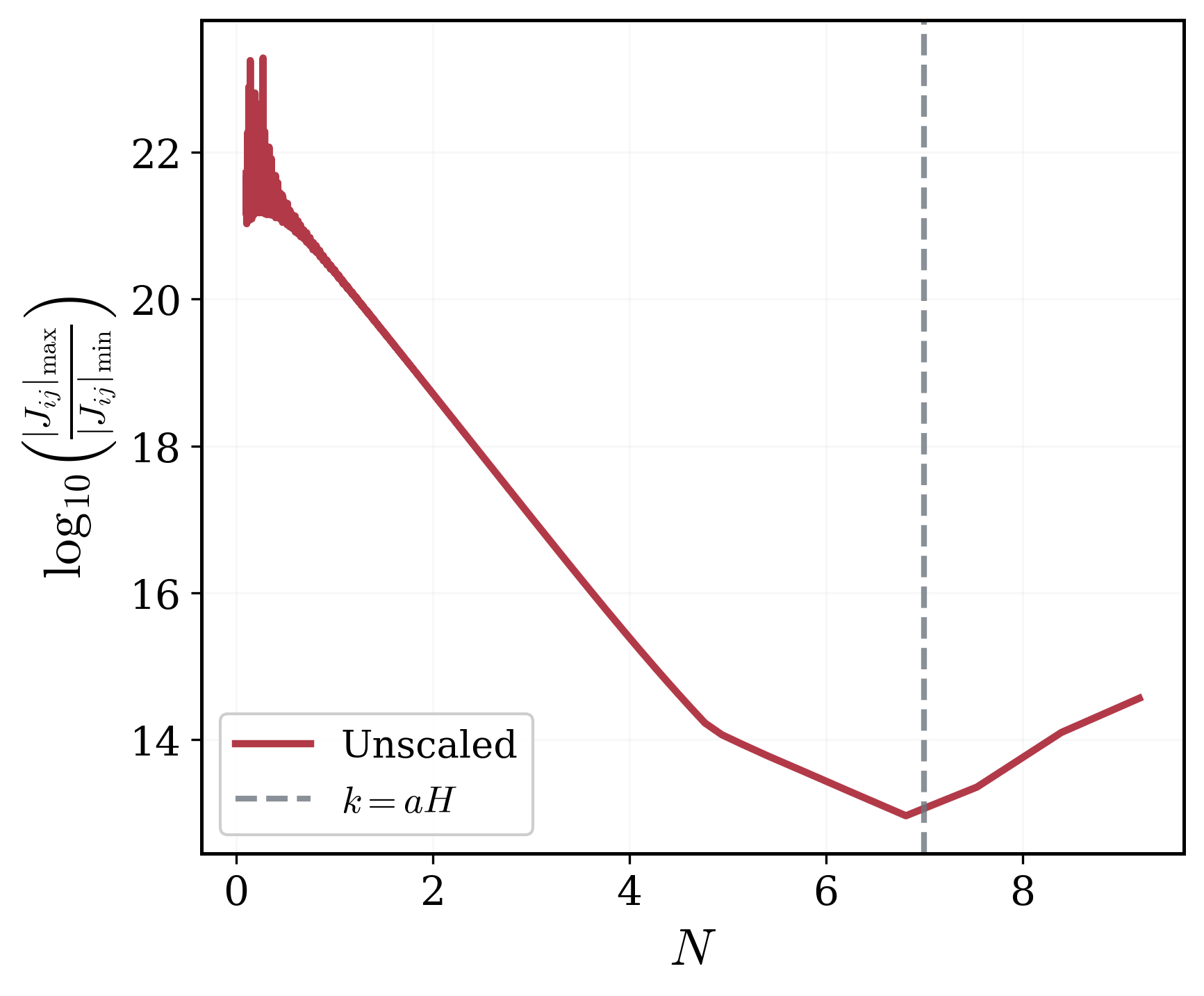}
\end{subfigure}

\caption{
Numerical behaviour of the unscaled deterministic evolution for the quartic potential model
$V(\phi)=V_0\phi^4/4$ with $\Upsilon\propto T^3$, using
$V_0=10^{-8}$, $g_*=106.75$ and $N_e=60$, assuming no inflaton thermalization ($n=0$) and neglecting the thermal noise contribution to the radiation perturbation equation.
\textbf{Left panel:} correction factor $G(Q)$ obtained using \texttt{WI2easy} and the unscaled deterministic implementation in \texttt{DSWIM}.
\textbf{Right panel:} evolution of the covariance matrix dynamic range (defined in~\cref{eq:dynamic_range}),
during the unscaled deterministic evolution. The vertical dashed line denotes horizon crossing ($k=aH$).
}
\label{fig:WDR_discrepancy}
\end{figure}

To quantify the severity of these hierarchies, we monitor the dynamic range of the covariance matrix during the deterministic evolution, defined as
\begin{equation}\label{eq:dynamic_range}
    \mathcal{D} = \log_{10}\left(\frac{|J_{ij}|_{\max}}{|J_{ij}|_{\min}}\right),
\end{equation}
where $|J_{ij}|_{\max}$ and $|J_{ij}|_{\min}$ denote the largest and smallest nonzero absolute values of the covariance matrix elements at a given efold time. Large values of this quantity indicate that different elements of the covariance matrix differ by many orders of magnitude, making the evolution increasingly susceptible to round-off errors and loss of numerical precision. These numerical effects become particularly evident in the weak dissipative regime, where the correction factor $G(Q)$ is expected to approach unity. In this regime, even small numerical inaccuracies become readily visible as artificial deviations from the expected behaviour. Figure~\ref{fig:WDR_discrepancy} illustrates this effect for a quartic potential model. The left panel shows numerical artifacts in the unscaled deterministic evolution of $G(Q)$ in the WDR, while the right panel shows that the covariance matrix develops a hierarchy spanning approximately $13$--$22$ orders of magnitude during the evolution. This behaviour is consistent with the interpretation that large hierarchies in the covariance matrix can adversely affect the numerical accuracy of the deterministic evolution.

\begin{figure}[htbp]
\centering
\begin{subfigure}[t]{0.49\linewidth}
    \centering
    \includegraphics[width=\linewidth]{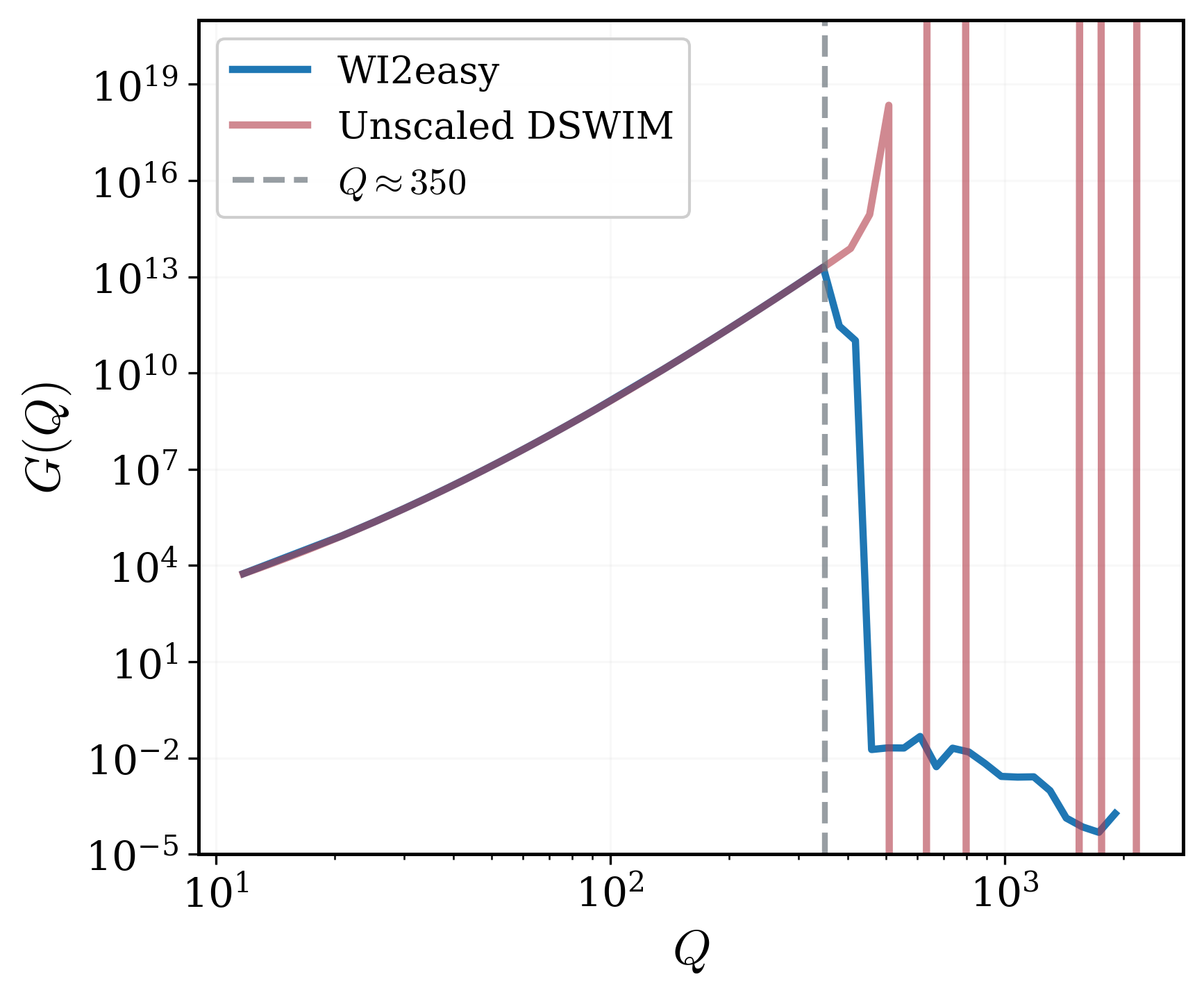}
\end{subfigure}
\hfill
\begin{subfigure}[t]{0.49\linewidth}
    \centering
    \includegraphics[width=\linewidth]{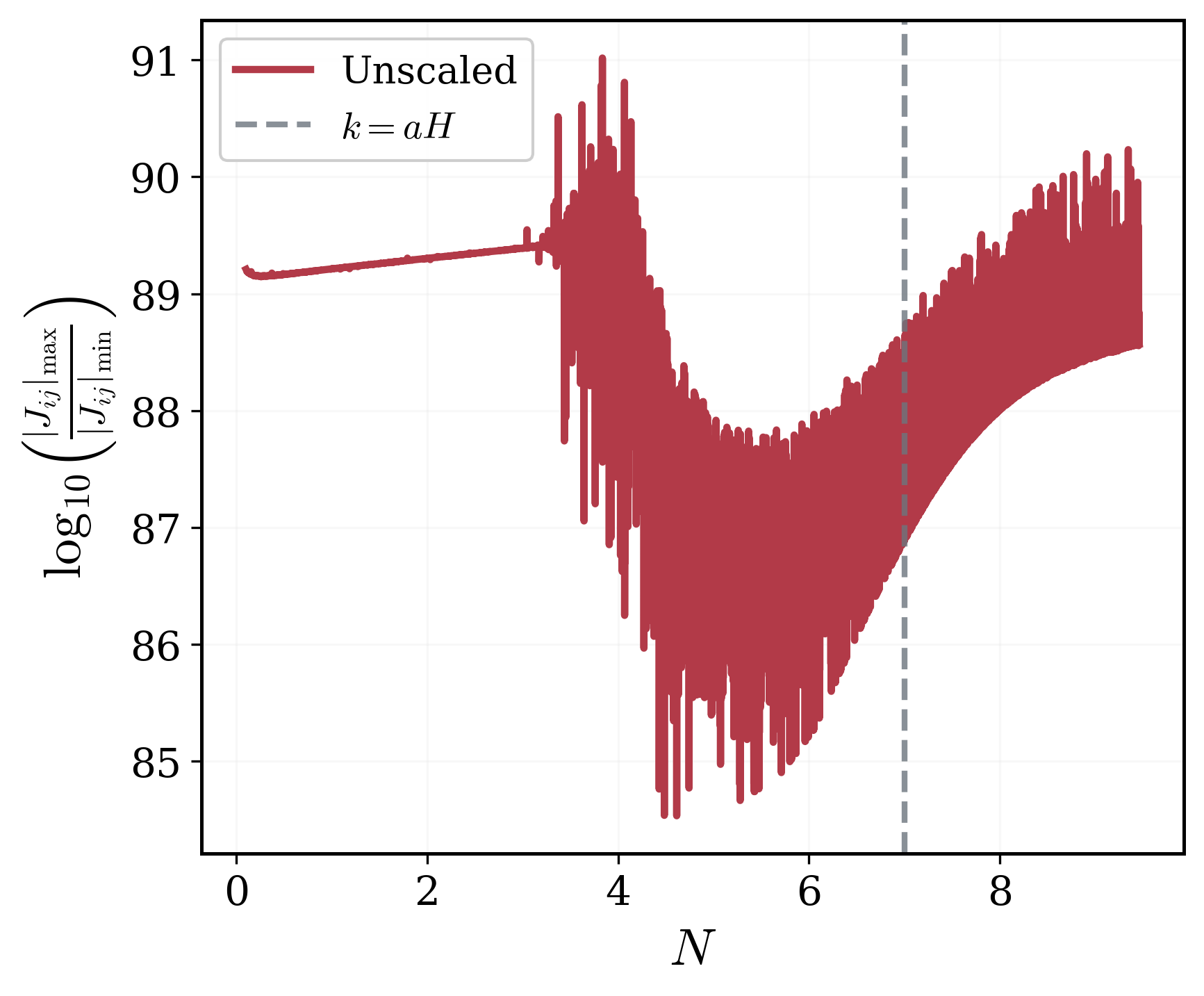}
\end{subfigure}
\caption{
Numerical behaviour of the unscaled deterministic evolution for the runaway exponential potential
$V(\phi)=V_0e^{-\alpha\phi^2}$ with $\Upsilon\propto T^3$, using
$\alpha=0.2$, $V_0=10^{-14}$, $g_*=106.75$ and $N_e=60$, assuming no inflaton thermalization ($n=0$) and neglecting the thermal noise contribution to the radiation perturbation equation.
\textbf{Left panel:} correction factor $G(Q)$ obtained using \texttt{WI2easy} and the unscaled deterministic implementation in \texttt{DSWIM}.
\textbf{Right panel:} evolution of the covariance matrix dynamic range 
(defined in~\cref{eq:dynamic_range}),
during the unscaled deterministic evolution. The vertical dashed line denotes horizon crossing ($k=aH$).
}
\label{fig:exp_pot_discrepancy}
\end{figure}

The impact of these hierarchies becomes even more pronounced at larger values of the dissipation ratio $Q$ for certain WI models. An example is shown in figure~\ref{fig:exp_pot_discrepancy} for the runaway exponential potential $V(\phi)=V_0 e^{-\alpha\phi^2}$, where the exponential suppression drives the potential energy density to extremely small values while the inflaton field amplitude continues to grow. In this case, the covariance matrix hierarchy approaches about $90$ orders of magnitude during the evolution, as shown in the right panel of figure~\ref{fig:exp_pot_discrepancy}. The resulting loss of numerical precision eventually leads to breakdown of the unscaled deterministic evolution. While \texttt{WI2easy} partially mitigates these effects through Mathematica's automatic integrator selection and stiffness handling, the underlying instability remains evident, indicating that the problem originates from the conditioning of the deterministic system itself rather than from a particular numerical implementation.

These numerical difficulties are not associated with the deterministic formalism itself, but rather with the conditioning of the correlation matrix evolution equations when evolved directly in terms of $\mathbf{J}$. To address this issue, we introduce in the following section a scaling framework based on a deterministic diagonal transformation matrix. The scaling transformation preserves the physical power spectrum exactly while substantially improving the numerical conditioning of the system.

\subsection{Scaling Matrix Framework}
\label{sec:S_framework}

To address the numerical difficulties discussed in the previous section, we introduce a scaling framework for the deterministic evolution equations. The framework is based on a scaling matrix $\mathbf{S}$ that normalizes the perturbation variables using appropriate powers of the Hubble parameter $H$, which sets the characteristic energy scale during inflation. The scaled state vector is defined as
\begin{equation}\label{eq:S_definition} \tilde{\mathbf{\Phi}} = \mathbf{S \Phi} = \begin{pmatrix}
\psi \\ 
\delta q_r/H \\ 
\delta \phi/H \\ 
\delta \rho_r/H^2\\ 
\delta \phi'/H 
\end{pmatrix}\, , 
\end{equation} 
where $\tilde{\mathbf{\Phi}}$ is the scaled state vector. The corresponding scaling matrix is therefore given by
\begin{equation}
\mathbf{S}=\mathrm{diag}\left(1,H^{-1},H^{-1},H^{-2},H^{-1}\right)\, .
\end{equation}
The powers of $H$ are chosen to normalize the perturbation variables by their characteristic scaling during inflation. The metric perturbation $\psi$ is dimensionless, while $\delta\phi$ and $\delta\phi'$ both have mass dimension $[M]$ and are therefore normalized by a single power of $H$. The radiation energy density perturbation $\delta\rho_r$ has mass dimension $[M]^4$; however, the background radiation density satisfies $\rho_r \propto H^2 M_{\rm Pl}^2$ through the Friedmann~\cref{eq:Friedmann}. After setting $M_{\rm Pl}=1$, both $\rho_r$ and its perturbation naturally scale as $H^2$, motivating the normalization $\delta\rho_r/H^2$. The radiation momentum perturbation $\delta q_r$ has mass dimension $[M]^3$. Rather than following directly from its mass dimension, the normalization $\delta q_r/H$ is motivated by the structure of the perturbation~\cref{eq:pert_rad,eq:pert_m}, where $\delta q_r$ repeatedly appears in the combination $\delta q_r/H$. This choice brings $\delta q_r$ to a scale comparable to the remaining perturbation variables and improves the conditioning of the deterministic system.The resulting transformation substantially reduces the spread of scales present in the perturbation system, preventing the loss of numerical precision.

It can be readily shown that this scaling transformation does not modify the comoving curvature perturbation provided that
\begin{equation}
\tilde{\mathbf{C}} = \mathbf{S}^{-1}\mathbf{C} \, ,
\end{equation}
the comoving curvature perturbation can then be written as
\begin{equation}
\mathcal{R} = \tilde{\mathbf{C}}^{T}\tilde{\mathbf{\Phi}} \, ,
\end{equation}
or equivalently,
\begin{equation}
\mathcal{R} = \mathbf{C}^{T}\mathbf{S}^{-1}\tilde{\mathbf{\Phi}} \, ,
\end{equation}
which is identical to~\cref{eq:matrx_R} using the scaling relation in~\cref{eq:S_definition}. The transformed deterministic system can be derived by differentiating~\cref{eq:S_definition} with respect to the number of e-folds $N$ and using the matrix Langevin~\cref{eq:matrix_langevin}, giving
\begin{equation}
\tilde{\mathbf{\Phi}}'= \left( \mathbf{S}'\mathbf{S}^{-1} + \mathbf{SAS}^{-1} \right) \tilde{\mathbf{\Phi}} + \mathbf{SB}_{T}\xi_T + \mathbf{SB}_{q}\xi_q \, .
\end{equation}
The transformed drift and diffusion matrices are therefore given by
\begin{equation}
\tilde{\mathbf{A}}=\mathbf{S}'\mathbf{S}^{-1}+\mathbf{SAS}^{-1}\, ,
\end{equation}
and using~\cref{eq:diffusion_matrix}
\begin{equation}
\tilde{\mathbf{D}}=\mathbf{SDS}^{T}\, .
\end{equation}
In the scaled framework, the evolved correlation matrix is the scaled correlation matrix $\tilde{\mathbf{J}}$, related to the original correlation matrix $\mathbf{J}$ through the transformation
\begin{equation}
\tilde{\mathbf{J}}=\mathbf{SJS}^{T}\, ,
\end{equation}
which follows directly from the definition of $\mathbf{J}$ in~\cref{eq:J_correlation}. The scaling transformation therefore keeps the elements of the evolved correlation matrix numerically comparable throughout the evolution. We implement this scaling framework alongside the stochastic formalism in \texttt{SWIM}, allowing the user to choose between stochastic and deterministic evolution approaches.

\section{Results and Comparisons}
\label{sec:Results}
In this section we present the results of the scaling matrix deterministic framework for solving WI scalar perturbations and compare it with the existing deterministic approach implementation by \texttt{WI2easy}. We also give an example in which the stochastic and deterministic versions of \texttt{SWIM} are compared with each other.

\subsection{Stabilization of Deterministic Evolution}

In the previous section, we showed that direct evolution of the unscaled deterministic system can lead to significant numerical instabilities for WI models exhibiting large hierarchies of scales in the perturbation evolution. We now demonstrate that the scaling framework introduced in \cref{sec:S_Formalism} substantially improves the numerical stability of the deterministic evolution while preserving agreement with the stochastic approach.

\begin{figure}[htbp]
\centering
\includegraphics[width=0.7\linewidth]{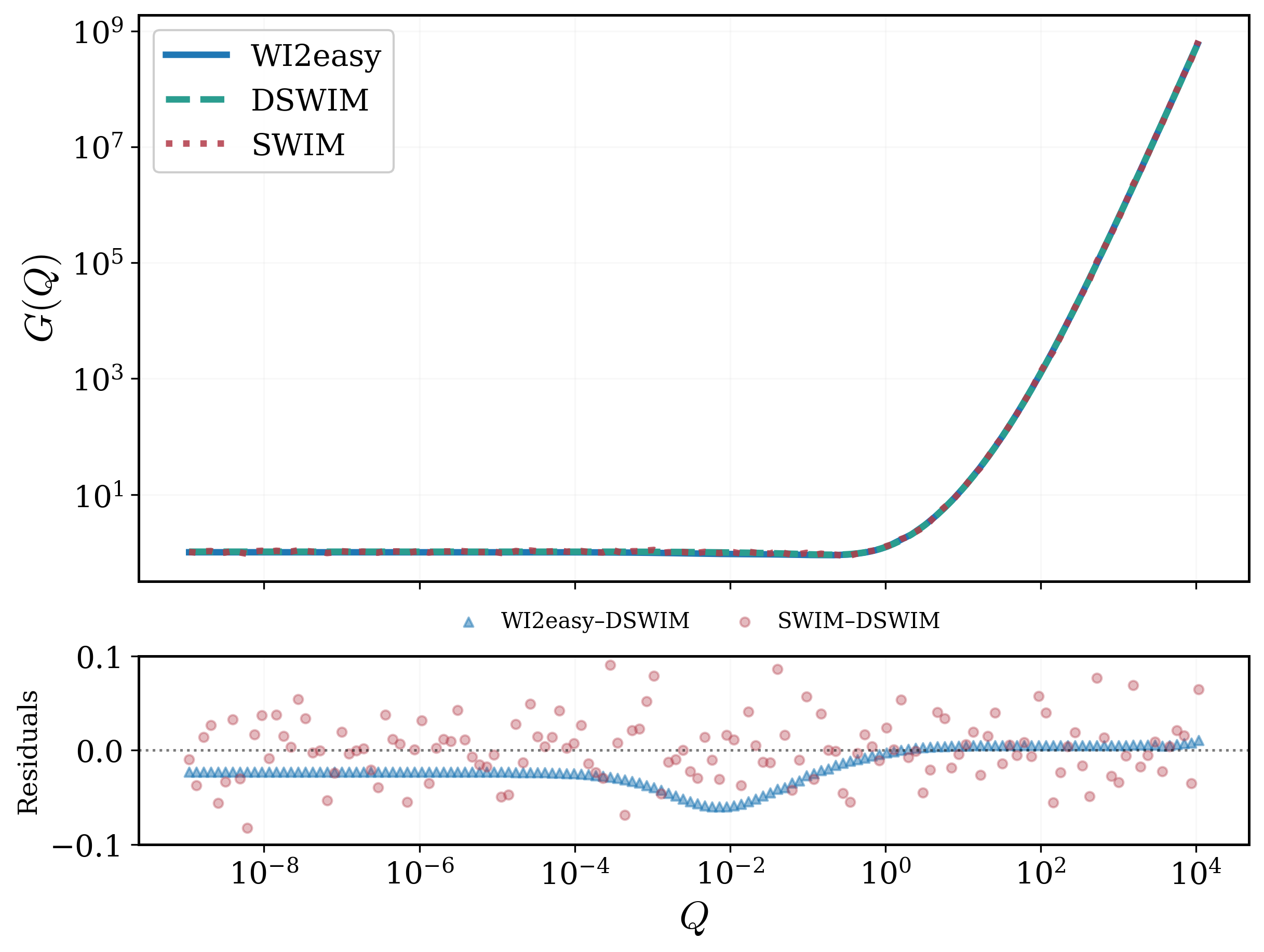}
\caption{
Comparison of the correction factor $G(Q)$ computed using \texttt{WI2easy}, the scaled deterministic implementation in \texttt{DSWIM}, and the stochastic implementation in \texttt{SWIM} for the quadratic warm inflation model with potential $V(\phi)=V_0\phi^2/2$ and dissipation coefficient $\Upsilon\propto T$. The model parameters are chosen as $V_0=10^{-14}$, $g_*=106.75$ and $N_e=60$, assuming no inflaton thermalization ($n=0$) and neglecting the thermal noise contribution to the radiation perturbation equation.
\textbf{Upper panel:} correction factor $G(Q)$ as a function of the dissipation ratio $Q$. The stochastic \texttt{SWIM} results are obtained using $2048$ realizations.
\textbf{Lower panel:} residuals relative to the scaled deterministic implementation in \texttt{DSWIM}, defined as $(G-G_{\mathrm{\texttt{DSWIM}}})/G_{\mathrm{\texttt{DSWIM}}}$.
}
\label{fig:quadratic_validation}
\end{figure}

Before considering WI models exhibiting severe hierarchies of scales, it is important to verify that the scaled deterministic implementation reproduces the results of existing stochastic and deterministic approaches in regimes where all methods remain numerically stable. Figure~\ref{fig:quadratic_validation} presents such a comparison for a representative quadratic WI model. The scaled deterministic implementation in \texttt{DSWIM} remains in good agreement with both stochastic \texttt{SWIM} and \texttt{WI2easy} over several orders of magnitude in the dissipation ratio, thereby validating the implementation of the deterministic formalism and the scaling framework. The lower panel shows the residuals with respect to \texttt{DSWIM}. The residuals between stochastic \texttt{SWIM} and deterministic \texttt{DSWIM} exhibit the expected statistical scatter arising from the finite number of stochastic realizations, while the agreement with \texttt{WI2easy} demonstrates consistency between independent deterministic implementations. A small but systematic discrepancy between \texttt{DSWIM} and \texttt{WI2easy} is nevertheless observed in the moderate dissipative regime, $Q\sim 10^{-3}-10^{-1}$. As discussed in appendix~\ref{appendix:quantum-noise}, this discrepancy arises from the different prescriptions adopted for incorporating the quantum contribution to the primordial fluctuations.

We first consider the quartic potential $V(\phi)=V_0\phi^4/4$ with dissipation coefficient $\Upsilon \propto T^3$. As shown in figure~\ref{fig:WDR_instability_solution}, the scaled deterministic implementation suppresses the numerical artifacts observed in the unscaled evolution and remains consistent with the stochastic result throughout the range of $Q$ considered. The right panel shows that the scaling transformation reduces the covariance matrix hierarchy by approximately $2$--$3$ orders of magnitude throughout the evolution, with the largest reduction occurring near horizon crossing. The same moderate-Q discrepancy with \texttt{WI2easy}, discussed above, is again visible in this model.

We next consider the runaway exponential potential $V(\phi)=V_0 e^{-\alpha\phi^2}$ with $\alpha=0.2$ and $\Upsilon \propto T^3$. This model exhibits a significantly larger hierarchy of scales due to the exponential suppression of the potential at large field values. As shown in figure~\ref{fig:Exp_instability_solution}, the unscaled deterministic implementation in \texttt{WI2easy} becomes numerically unstable around $Q\approx350$. In contrast, the scaling framework implemented in \texttt{DSWIM} extends the stable evolution to significantly larger values of $Q$ and remains consistent with the stochastic result over nearly an order of magnitude larger dissipation ratios. Compared to the unscaled evolution shown in figure~\ref{fig:exp_pot_discrepancy}, the scaling transformation reduces the covariance matrix hierarchy by approximately $40$--$50$ orders of magnitude throughout the evolution, as shown in the right panel of figure~\ref{fig:Exp_instability_solution}. This substantial reduction in the hierarchy is accompanied by a corresponding extension of the stable evolution range of the deterministic system. The eventual breakdown of both stochastic and deterministic \texttt{SWIM} around $Q\approx2000$ occurs when the exponentially suppressed potential drives several background and perturbation quantities towards machine precision.

\begin{figure}[htbp]
    \centering
    \begin{subfigure}[t]{0.49\linewidth}
    \centering
    \includegraphics[width=\linewidth]{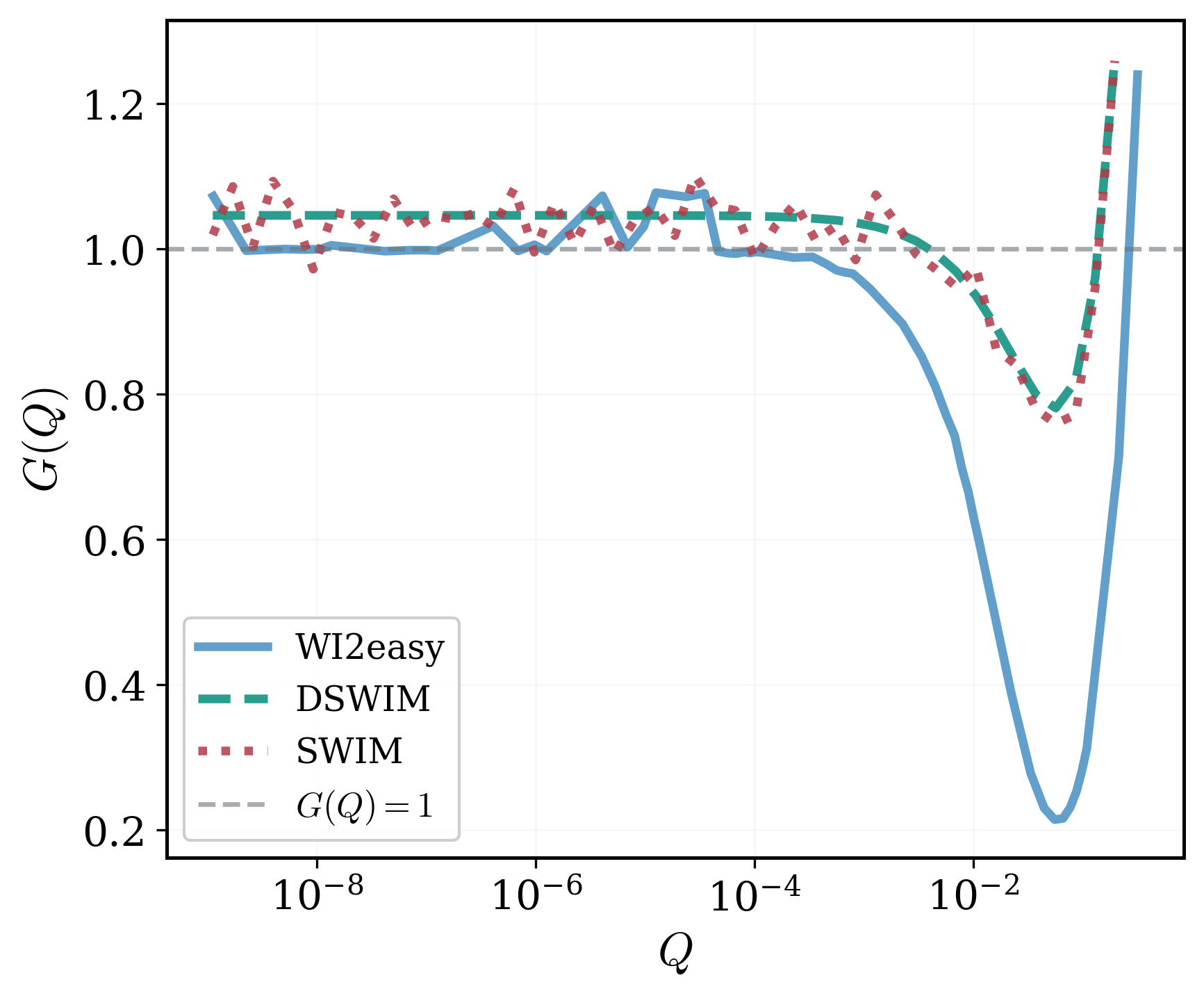}
    \end{subfigure}
    \hfill
    \begin{subfigure}[t]{0.49\linewidth}
        \centering
        \includegraphics[width=\linewidth]{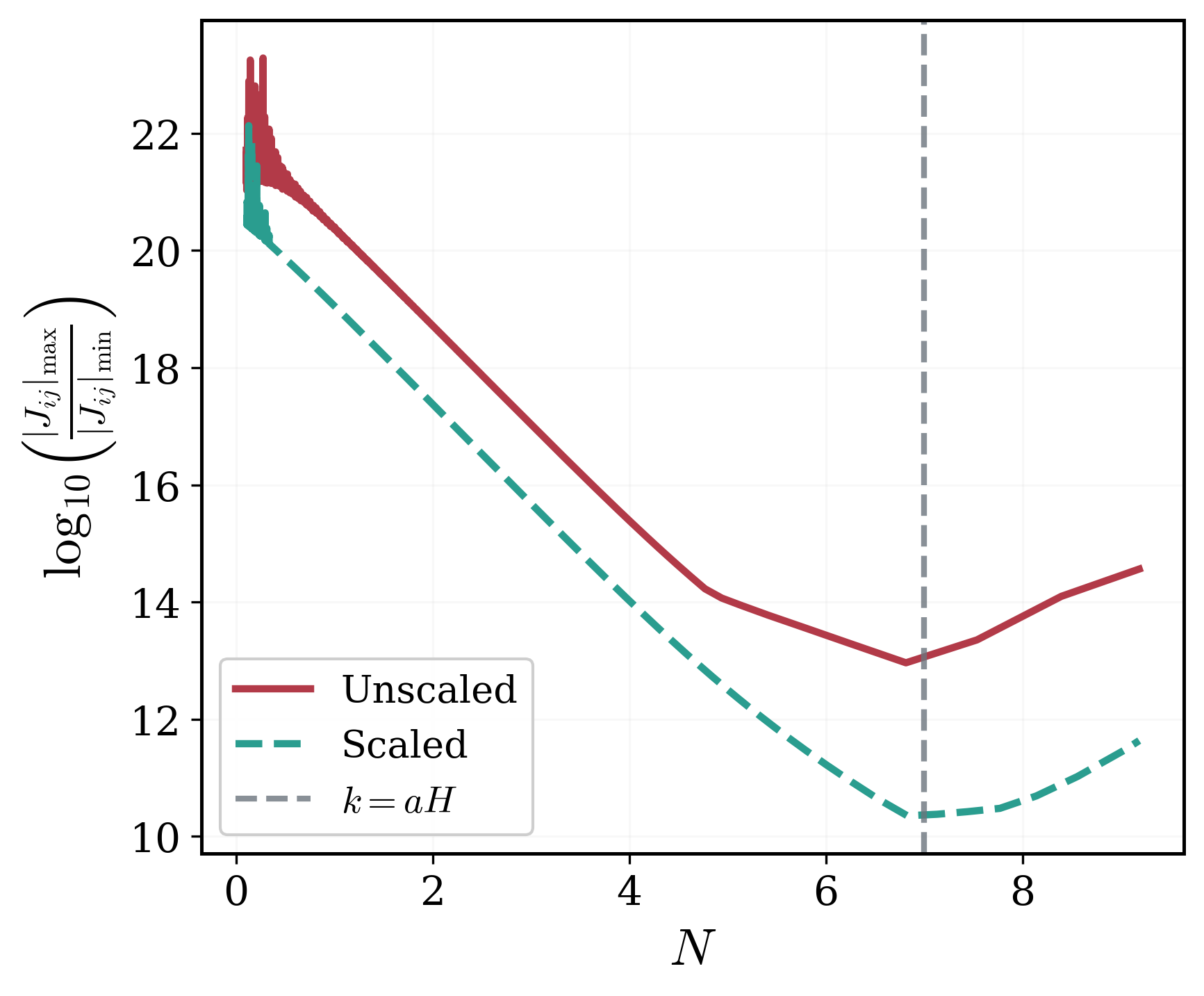}
    \end{subfigure}
    \caption{
Comparison of the correction factor $G(Q)$ obtained using \texttt{WI2easy}, the stochastic implementation in \texttt{SWIM}, and the scaled deterministic implementation in \texttt{DSWIM} for the same WI model and parameters as figure~\ref{fig:WDR_discrepancy}.
\textbf{Left panel:} correction factor $G(Q)$ as a function of the dissipation ratio $Q$.
\textbf{Right panel:} evolution of the covariance matrix dynamic range (defined in~\cref{eq:dynamic_range}), for the scaled and unscaled deterministic evolution in \texttt{DSWIM}.
}
\label{fig:WDR_instability_solution}
    \label{fig:WDR_instability_solution}
\end{figure}

\begin{figure}[htbp]
    \centering
    \begin{subfigure}[t]{0.49\linewidth}
    \centering
    \includegraphics[width=\linewidth]{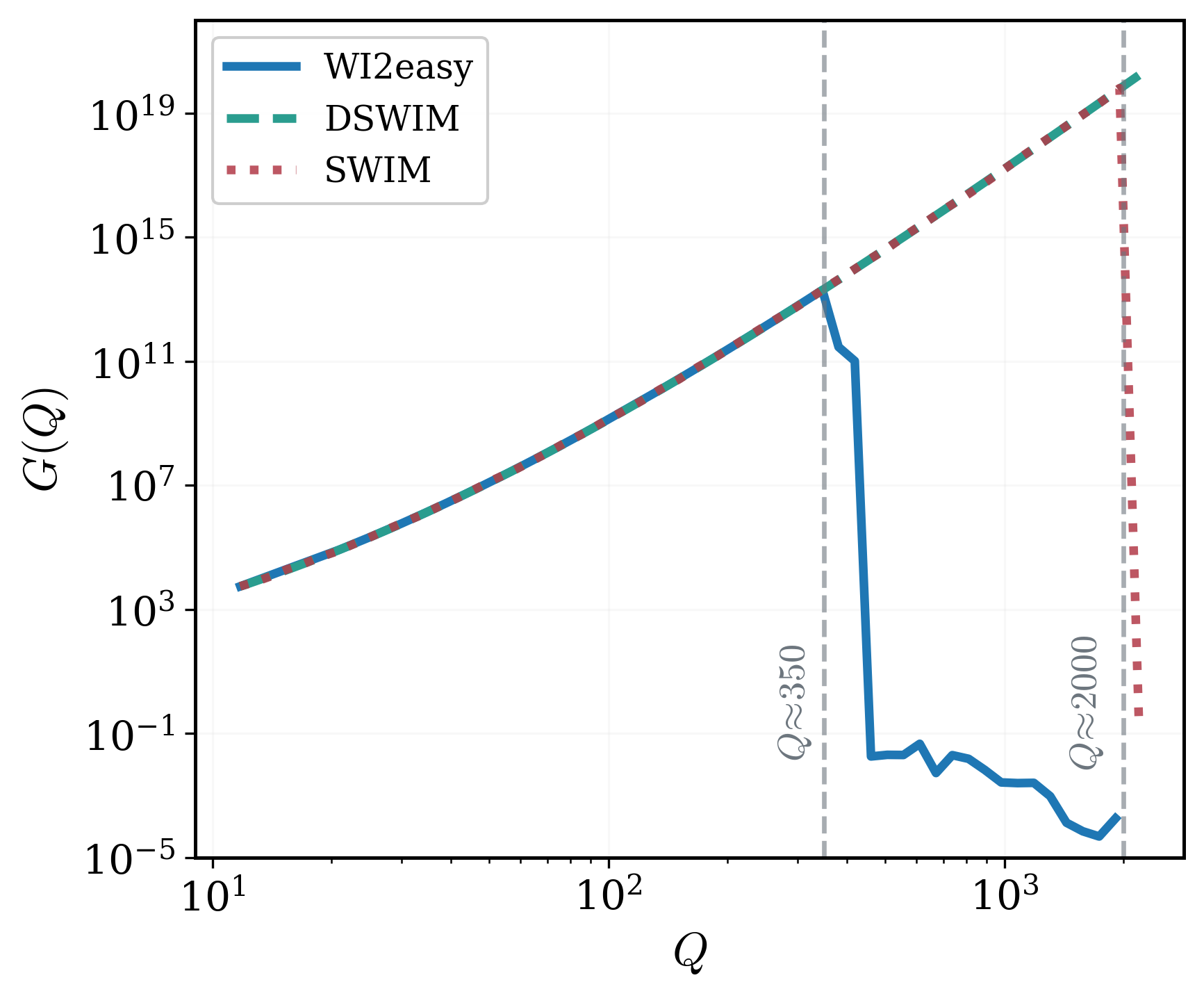}
    \end{subfigure}
    \hfill
    \begin{subfigure}[t]{0.49\linewidth}
        \centering
        \includegraphics[width=\linewidth]{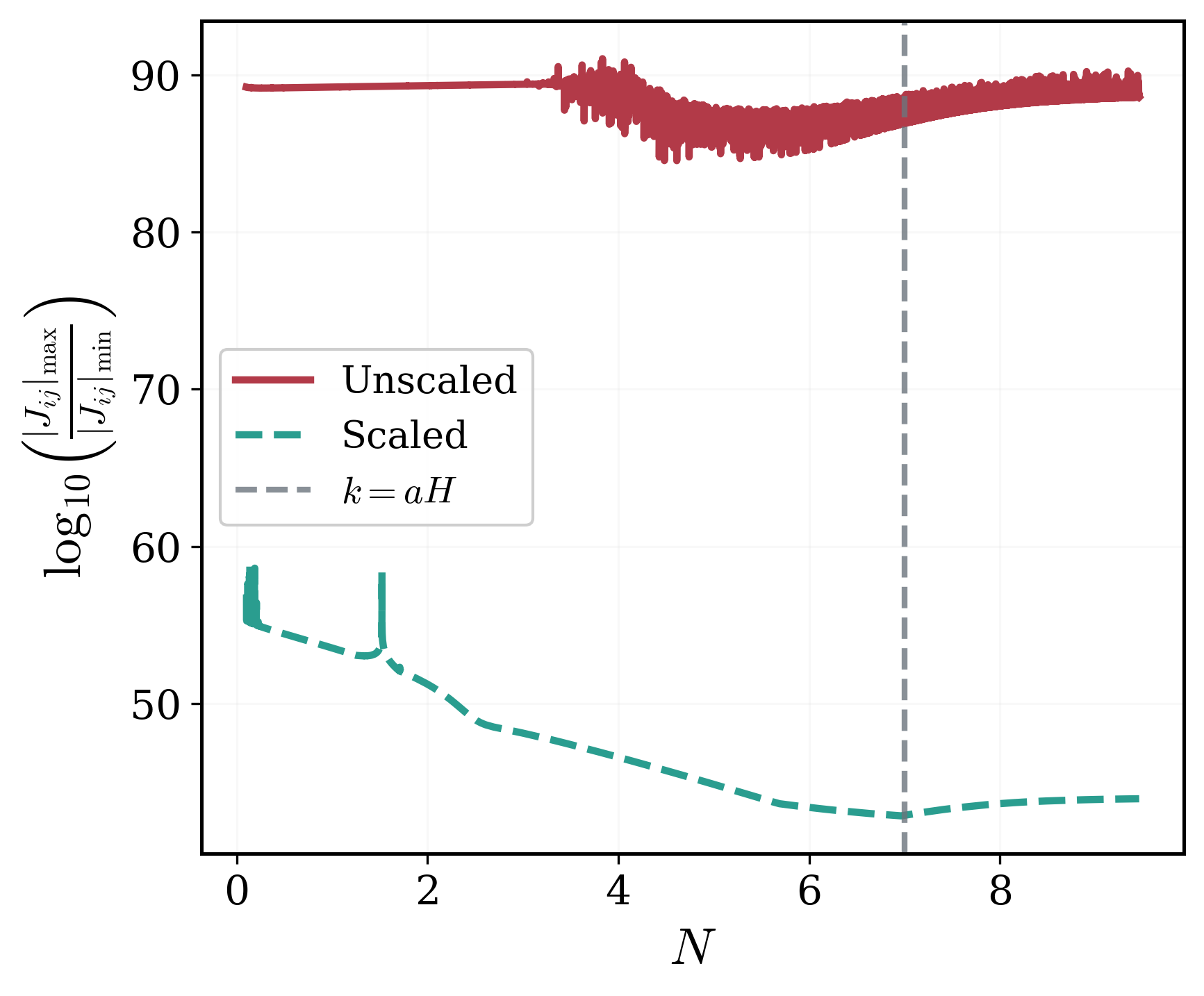}
    \end{subfigure}
    \caption{
Comparison of the correction factor $G(Q)$ obtained using \texttt{WI2easy}, the scaled deterministic implementation in \texttt{DSWIM}, and the stochastic implementation in \texttt{SWIM} for the runaway exponential potential discussed in figure~\ref{fig:exp_pot_discrepancy}.
\textbf{Left panel:} correction factor $G(Q)$ as a function of the dissipation ratio $Q$. The vertical dashed lines indicate the approximate onset of numerical instability in \texttt{WI2easy} ($Q\approx350$) and \texttt{SWIM} ($Q\approx2000$).
\textbf{Right panel:} evolution of the covariance matrix dynamic range (defined in~\cref{eq:dynamic_range}), for the scaled and unscaled deterministic evolution in \texttt{DSWIM}.
}
    \label{fig:Exp_instability_solution}
\end{figure}

Taken together, these results demonstrate that the scaling transformation substantially improves the numerical conditioning of the deterministic system while preserving the physical predictions of the perturbation evolution exactly. The scaled deterministic formulation remains consistent with existing approaches in numerically well-behaved regimes, suppresses numerical artifacts in the WDR, and significantly extends the stable evolution range in the SDR.

\subsection{Noise Correlations in the Deterministic Formalism}

\begin{figure}[htbp]
    \centering
    \includegraphics[width=0.7\linewidth]{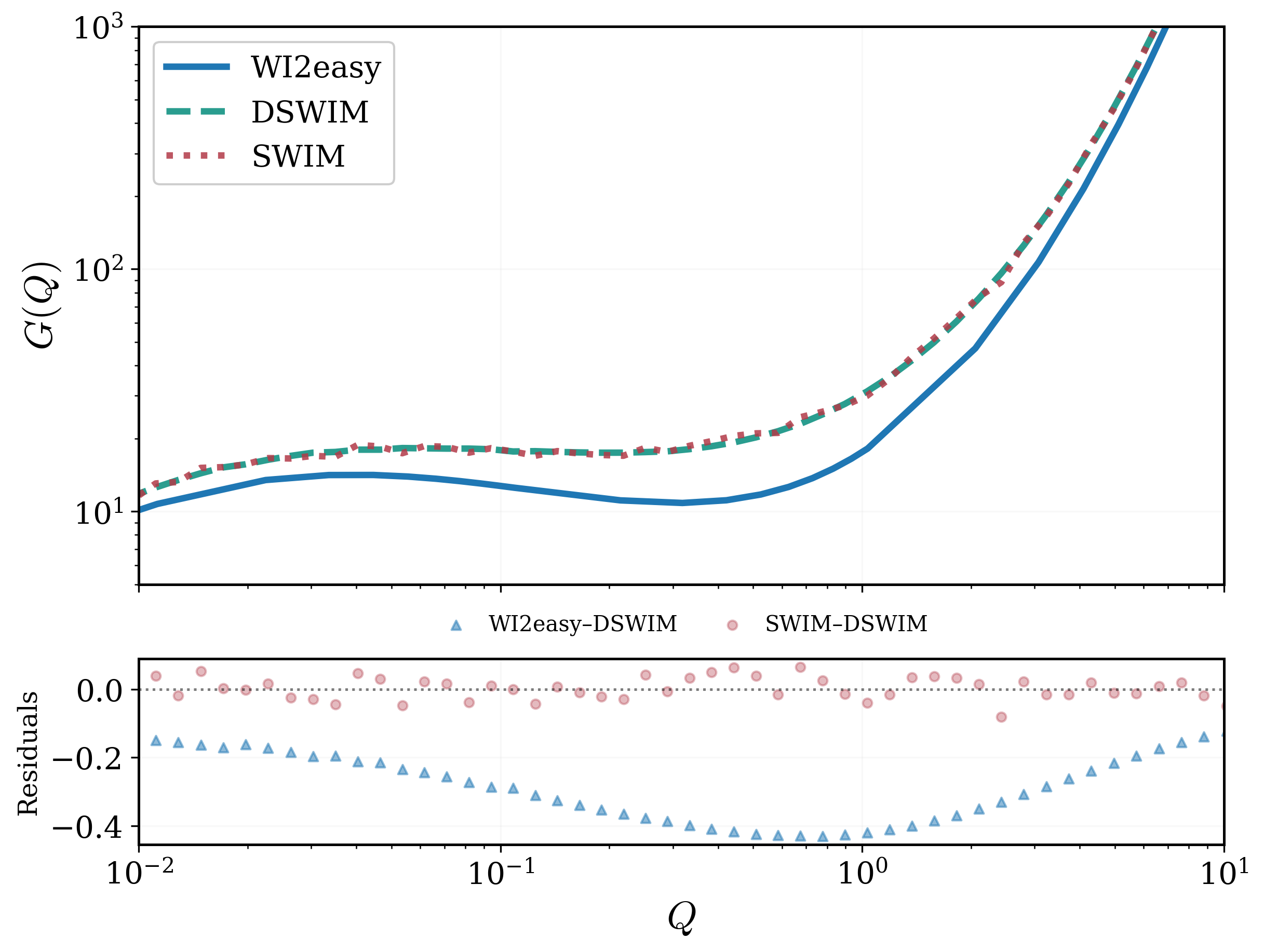}
    \caption{
Comparison of the correction factor $G(Q)$ obtained using \texttt{WI2easy}, stochastic \texttt{SWIM}, and deterministic \texttt{DSWIM} for the quartic potential $V(\phi)=V_0\phi^4/4$ with dissipation coefficient $\Upsilon \propto T^3$. The parameters are chosen as $V_0=10^{-14}$, $g_*=106.75$ and $N_e=60$, including the thermal noise contribution in the radiation perturbation equation while assuming no inflaton thermalization ($n=0$). The upper panel shows the correction factor $G(Q)$ as a function of the dissipation ratio $Q$, while the lower panel shows residuals relative to \texttt{DSWIM}.
}
    \label{fig:moderateQ_radnoise}
\end{figure}

The equivalence between the stochastic and deterministic approaches was previously demonstrated in refs.~\cite{Ballesteros:2022hjk,Ballesteros:2023dno}. More recently, ref.~\cite{Kumar:2026mvz} compared the stochastic implementation in \texttt{SWIM} with the deterministic implementation in \texttt{WI2easy} through the correction factor $G(Q)$ for several representative WI models. However, when the thermal noise term is included in the radiation perturbation equation, a systematic discrepancy appears in the moderate dissipative regime $0.01\lesssim Q \lesssim 10$ for all the exemplary models studied in \cite{Kumar:2026mvz} (see figures~3 and 4 therein). An explicit example of this discrepancy is shown in figure~\ref{fig:moderateQ_radnoise}.

This discrepancy is now understood to originate from the treatment of the thermal noise terms in \texttt{WI2easy}. In the deterministic formulation implemented there, the thermal noise contributions appearing in the inflaton and radiation perturbation equations are assumed to be independent. However, since the thermal noise terms $\xi_T$ appearing in the inflaton and radiation perturbation equations arise from energy-momentum conservation, they must be correlated and therefore generate non-vanishing off-diagonal contributions to the diffusion matrix $\mathbf{D}$.

In the deterministic framework implemented in \texttt{DSWIM}, these cross correlations are explicitly incorporated through the construction of the diffusion matrix in~\cref{eq:diffusion_matrix}. The corresponding off-diagonal terms can be seen explicitly in the appendix~\ref{eq:D-matrix_explicit}. As shown in figure~\ref{fig:moderateQ_radnoise}, the deterministic implementation in \texttt{DSWIM} remains in excellent agreement with the stochastic result in the moderate dissipative regime once these correlations are properly accounted for.

It is important to emphasize that this issue is specific to the publicly available implementation of \texttt{WI2easy} rather than to the deterministic formalism itself. The deterministic formulation developed in refs.~\cite{Ballesteros:2022hjk,Ballesteros:2023dno} already incorporates the correlated thermal fluctuations through a non-diagonal diffusion matrix. One can verify that, up to the treatment of the quantum contribution and the ordering of the perturbation variables, the diffusion matrix employed in \texttt{DSWIM} is equivalent to the noise correlation matrix $\mathbf{BB}^{\mathrm T}$ given in eq.~(3.15) of ref.~\cite{Ballesteros:2023dno}, where the noise matrix $\mathbf{B}$ is defined in appendix A.3 of the same reference.

\subsection{Computational Performance and Practical Advantages}

The stochastic calculation of the WI power spectrum requires ensemble averaging over a large number of realizations, making parameter space exploration computationally expensive. In contrast, the deterministic formalism evolves the correlation matrix directly through the coupled ODE system in~\cref{eq:FP_ode}, requiring only a single deterministic integration for each comoving mode $k$. This leads to a substantial improvement in computational efficiency.

As demonstrated in the previous sections, direct evolution of the unscaled deterministic system can suffer from severe numerical instabilities for WI models exhibiting large hierarchies of scales. The scaling framework implemented in \texttt{DSWIM} restores numerical stability while preserving agreement with the stochastic formulation, thereby combining the computational advantages of the deterministic formalism with the robustness of the stochastic approach.

\begin{table}[htbp]
\centering
\begin{tabular}{lccc}
\hline
Model & Code/Method & \texttt{Find\_ICs} [s] & \texttt{Find\_GQ} [s] \\
\hline
Quartic $V(\phi)=V_0\phi^4/4$
& \texttt{WI2easy} & 12.78 & 123.3 \\
& \texttt{SWIM} (Stochastic) & 9.44 & 116.49 \\
& \texttt{DSWIM} (Deterministic) & - & 1.56 \\
\hline
Runaway $V(\phi)=V_0 e^{-\alpha\phi^2}$ & \texttt{WI2easy} & 46.92  &  233.57\\
& \texttt{SWIM} (Stochastic) & 599.81 & 913.86 \\
& \texttt{DSWIM} (Deterministic) & - &  8.18 \\
\hline
Quadratic $V(\phi)=V_0\phi^2/2$
& \texttt{WI2easy} & 66.98  &  750.6 \\
& \texttt{SWIM} (Stochastic) &  1187.41 & 1076.52 \\
& \texttt{DSWIM} (Deterministic) & - &  22.66\\
\hline
\end{tabular}
\caption{Runtime comparison between \texttt{WI2easy} and stochastic/deterministic \texttt{SWIM} for representative WI models. The quantity \texttt{Find\_ICs} denotes the runtime required for obtaining the background initial conditions, while \texttt{Find\_GQ} denotes the runtime required for computing the correction factor $G(Q)$. Since the background evolution is deterministic, the \texttt{Find\_ICs} runtime for deterministic \texttt{SWIM} is identical to the stochastic implementation and is therefore not quoted separately. The deterministic implementation in \texttt{SWIM} achieves a substantial reduction in runtime relative to both the stochastic implementation and the deterministic implementation in \texttt{WI2easy}.}
\label{tab:runtime_comparison}
\end{table}

In table~\ref{tab:runtime_comparison}, we compare the runtimes of stochastic and deterministic \texttt{SWIM} for representative WI models alongside \texttt{WI2easy}.\footnote{Benchmarks were performed on a ThinkPad P14s Gen 6 running Ubuntu 24.04 (Linux kernel 6.14.11), equipped with an AMD Ryzen AI 7 PRO 350 processor (16 threads) and 32 GB RAM (27 GB usable). The software environment consisted of Mathematica 14.3, Python 3.14.2, \texttt{g++} 15.2, and Boost 1.91.} For the quartic potential model, \texttt{DSWIM} reduces the runtime of the $G(Q)$ computation from $116.49\,\mathrm{s}$ in the stochastic implementation to $1.56\,\mathrm{s}$, corresponding to a speedup of approximately two orders of magnitude. Similarly, for the runaway exponential potential, the runtime is reduced from $913.86\,\mathrm{s}$ to $8.18\,\mathrm{s}$, while for the quadratic potential it decreases from $1076.52\,\mathrm{s}$ to $22.66\,\mathrm{s}$.\footnote{The \texttt{Find\_ICs} stage of \texttt{SWIM} is often computationally slower than that of \texttt{WI2easy}. In \texttt{SWIM}, the initial conditions are obtained using a brute-force optimization procedure implemented using \texttt{SciPy}. While this approach is robust and applicable to a broad class of WI models, it requires multiple background integrations to identify initial conditions that yield the desired number of e-folds. Furthermore, as $Q$ increases towards the SDR, the background equations become increasingly stiff, causing the adaptive integrator to adopt smaller step sizes and thereby increasing the computational cost. Consequently, \texttt{SWIM} is comparatively faster in the WDR, while its runtime grows more rapidly in the SDR. In contrast, \texttt{WI2easy} employs a dedicated root-finding algorithm for the initial inflaton amplitude, typically requiring fewer background evaluations per initial-condition search.} Despite the additional matrix operations introduced by the scaling framework at each integration step, the stabilized deterministic implementation in \texttt{DSWIM} remains substantially faster than both the stochastic implementation and the deterministic implementation in \texttt{WI2easy}, while simultaneously improving numerical robustness for WI models exhibiting large hierarchies of scales in their perturbation evolution.

The scaled deterministic implementation in \texttt{DSWIM} is directly compatible with the existing \texttt{Cobaya} interface developed for stochastic \texttt{SWIM}, making it suitable for efficient parameter inference studies of WI models.

\section{Discussion and Conclusion}
In this work, we showed that although the deterministic formalism for WI perturbations is computationally more efficient than the stochastic approach, direct evolution of the unscaled deterministic system can become numerically ill-conditioned for WI models exhibiting large hierarchies of scales in the perturbation evolution. This leads to loss of numerical accuracy, artificial deviations in the correction factor $G(Q)$ and eventual breakdown of the deterministic evolution for sufficiently extreme models.

To address this issue, we introduced a scaling matrix framework for the deterministic evolution equations in which the perturbation variables are rescaled by appropriate powers of the Hubble parameter $H$. The scaling transformation is physically motivated by the effective scaling behaviour of the perturbation variables and substantially reduces the dynamic range of the deterministic system by keeping the scaled perturbation variables numerically comparable throughout the evolution. We showed that this scaling framework preserves the comoving curvature perturbation and consequently leaves the primordial scalar power spectrum invariant.

Using representative WI models, we demonstrated that the stabilized deterministic implementation in \texttt{DSWIM} substantially suppresses the numerical artifacts present in the unscaled deterministic evolution and significantly extends the stable evolution range for models exhibiting severe hierarchies of scales. We further showed that the deterministic implementation in \texttt{DSWIM} remains in excellent agreement with the stochastic formulation while simultaneously achieving a substantial reduction in computational cost relative to the stochastic approach.

We also discussed the role of correlated thermal noise contributions in the deterministic formalism. Since the thermal noise terms appearing in the inflaton and radiation perturbation equations arise from energy-momentum conservation, they must be correlated and therefore generate non-vanishing off-diagonal contributions to the diffusion matrix. Incorporating these correlations in deterministic \texttt{DSWIM} resolves the systematic discrepancy observed between the stochastic and deterministic approaches in the moderate dissipative regime.

The stabilized deterministic implementation developed in this work is directly compatible with the existing \texttt{Cobaya} interface in \texttt{SWIM}, making it suitable for efficient parameter inference studies of WI models. To the best of our knowledge, \texttt{DSWIM} is currently the only publicly available WI perturbation solver providing a numerically stabilized deterministic formalism directly interfaced with \texttt{Cobaya}.
\label{sec:Discussion}

\appendix
\section{Deterministic System Matrices}
\label{appendix:matrices}
In this section we present the explicit elements of the unscaled deterministic system of solving WI perturbations.
The elements of the drift matrix $\mathbf{A}$ can be obtained by comparing the system of \cref{eq:pert_KG,eq:pert_rad,eq:pert_mtm,eq:pert_m} and the matrix form of langevin~\cref{eq:matrix_langevin}. It is a $5 \times 5$ real valued non-symmetric matrix. The explicit elements can be read off as,
\begin{equation}\label{eq:A-matrix_explicit}
    \begin{aligned}
    &A_{00} = -1 , \quad A_{01} = -\frac{1}{2H} , \quad A_{02} = \frac{1}{2}\phi', \quad A_{03} = A_{04} = 0,
    \\[6pt]
    &A_{10} = -\frac{4 \rho_{r}}{3H}, \quad A_{11} = -3, \quad A_{12} = -\Upsilon \phi', \quad A_{13} = -\frac{1}{3H}, \quad A_{14} = 0, \\[6pt]
    &A_{20} = A_{21}= A_{22}=A_{23}=0, \quad A_{24} = 1, \\[6pt]
    &A_{30} = -\Upsilon H \phi'^2 -4 \rho_r, \quad A_{31} = \frac{k^2}{a^2 H} - \frac{2 \rho_r }{H}, \quad A_{32} = 2\rho_r  \phi' + \Upsilon_{,\phi}H\phi'^2, \\[6pt]
    &A_{33} = -4 + \frac{\Upsilon_{,T}H\phi'^2 T}{4 \rho_r }, \quad A_{34} = 2\Upsilon H \phi'\\[6pt]
    &A_{40} = -\frac{\Upsilon \phi'}{H} - \frac{2V_{,\phi}}{H^2}-4\phi', \quad A_{41} = -\frac{2\phi'}{H}, \quad A_{42} = -\frac{k^2}{a^2 H^2} -\frac{V_{,\phi\phi}}{H^2}-\frac{\Upsilon_{,\phi}\phi'}{H} + 2 \phi'^2, \\[6pt]
    &A_{43} = -\frac{\Upsilon_{,T}T\phi'}{4 H \rho_r}, \quad A_{44} = -3 -\frac{\Upsilon}{H} - \frac{H'}{H}\, .
    \end{aligned}
\end{equation}
The diffusion matrix $\mathbf{D}$ can be constructed by the noise column vectors $\mathbf{B}_T$ and $\mathbf{B}_q$

\begin{align}
    \mathbf{B}_T = \begin{pmatrix}
0 \\
0 \\
0 \\
-H^2 \phi' n_{T} \\
n_{T}
\end{pmatrix},
\qquad
\mathbf{B}_q = \begin{pmatrix}
0 \\
0 \\
0 \\
0 \\
n_q
\end{pmatrix}\, ,
\end{align}
where $n_T = \sqrt{ \frac{2\Upsilon T}{a^3H^3}}$ and $n_q = ( 9H + 4\pi\Upsilon )^{1/4} \sqrt{\frac{ 1 + 2n }{\pi a^3H^{3/2}}}$    are the thermal and quantum noises respectively. Then from~\cref{eq:diffusion_matrix} the drift matrix can be obtained explicitly

\begin{equation}\label{eq:D-matrix_explicit}
\begin{aligned}
    &D_{00} = D_{01} = D_{02} = D_{03} = D_{04} =0, \\[6pt]
    &D_{10} = D_{11} = D_{12} = D_{13} = D_{14} =0, \\[6pt]
    &D_{20} = D_{21} = D_{22} = D_{23} = D_{24} =0, \\[6pt]
    &D_{30} = D_{31} = D_{32} = 0, \quad  D_{33} = H^4\phi'^2 n_{T}^2, \quad D_{34} =-H^2 \phi' n_{T}^2, \\[6pt]
    &D_{40} = D_{41} = D_{42} = 0, \quad  D_{43} =-H^2 \phi' n_{T}^2, \quad  D_{44} =n_{T}^2 + n_{q}^2, \\[6pt]
\end{aligned}
\end{equation}
which is a real valued symmetric matrix. Finally the column vector $\mathbf{C}$ for computing comoving curvature perturbation can be derived using the ~\cref{eq:comoving_R,eq:matrx_R}
\begin{equation}\label{eq:C-matrix_explicit}
    C = \begin{pmatrix}
-1 \\[2pt]
\frac{H}{\bar{\rho}+\bar{p}} \\[6pt]
-\frac{H^2\phi'}{\bar{\rho}+\bar{p}}\\[2pt]
0 \\[2pt]
0 \\[2pt]
\end{pmatrix} \, .
\end{equation}

We implement these matrices along with the scaling matrix $\mathbf{S}$ to implement the framework presented in the section~\ref{sec:S_framework} in \texttt{SWIM} as the deterministic mode.

\section{Using the Deterministic Formalism in \texttt{SWIM}}

The scaled deterministic formalism implemented in this work can be enabled in \texttt{SWIM} by setting the parameter \texttt{want\_FP = 1} in the appropriate Python driver scripts for example, in \texttt{find\_GQ.py} of \texttt{GQ\_Calculator} submodule. This activates the deterministic Fokker-Planck evolution of the correlation matrix instead of the stochastic realization based evolution. Detailed instructions for using the deterministic mode are available in the documentation: \url{https://swim.readthedocs.io/en/latest/}.

Since the background evolution remains deterministic in both approaches, the same background initial conditions obtained through \texttt{find\_ICs.py} are used for both stochastic and deterministic evolution modes. The deterministic implementation is directly compatible with the existing \texttt{Cobaya} interface in \texttt{SWIM}, allowing efficient parameter inference studies of WI models. All other functionalities and usage of \texttt{SWIM} remain unchanged. For details of the code flow, we refer the reader to figures~2, 5, and 7 of ref.~\cite{Kumar:2026mvz}, while installation and usage instructions can be found in appendices A--D of ref.~\cite{Kumar:2026mvz}.

\section{Differences in the Treatment of the Quantum Contribution}
\label{appendix:quantum-noise}

The quantum contribution to the primordial fluctuations can be incorporated using different prescriptions, all of which are expected to reproduce the correct cold inflation limit. Existing implementations differ primarily in whether this contribution is generated through an explicit quantum stochastic source or through the homogeneous solution evolved from Bunch--Davies initial conditions. These choices are independent of whether the perturbation equations are solved using stochastic realizations or through the deterministic covariance matrix evolution. For completeness, we summarize the prescriptions adopted in the literature in table~\ref{tab:quantum_noise}. Throughout this appendix, $n=0$ denotes non-thermalized inflaton fluctuations, while $n=n_{\rm BE}$ denotes thermalized inflaton fluctuations following a Bose--Einstein distribution.

\begin{table}[ht]
\centering
\small
\renewcommand{\arraystretch}{1.3}
\begin{tabular}{p{2.5cm}p{2.6cm}p{3.4cm}p{5.0cm}}
\hline
Implementation & Quantum stochastic source & Initial conditions & Quantum prescription \\
\hline

Prescription in~\cite{Ballesteros:2022hjk,Ballesteros:2023dno} & No &
Bunch--Davies &
Quantum contribution obtained from the homogeneous solution. The homogeneous contribution is finally multiplied by the factor $(1+2n)$. 
\\[2mm]

\texttt{WI2easy} & \begin{tabular}[t]{@{}l@{}}
$n=0:$ No\\
$n=n_{\rm BE}:$ Yes
\end{tabular} & \begin{tabular}[t]{@{}l@{}}
$n=0:$ Bunch--Davies\\
$n=n_{\rm BE}:$ Vanishing
\end{tabular} & Hybrid prescription. \\[7mm]

\texttt{SWIM}/\texttt{DSWIM} & Yes & Vanishing &
Quantum contribution generated entirely through the quantum stochastic source.\\

\hline
\end{tabular}
\caption{Comparison of the prescriptions adopted in the literature for incorporating the quantum contribution to the primordial fluctuations.}
\label{tab:quantum_noise}
\end{table}

The prescription adopted by \texttt{WI2easy} differs from both the original deterministic formulation of refs.~\cite{Ballesteros:2022hjk,Ballesteros:2023dno} and the implementation adopted in \texttt{SWIM}/\texttt{DSWIM}. In the former, the quantum contribution is always obtained by evolving the homogeneous solution with Bunch--Davies initial conditions. In contrast, \texttt{WI2easy} employs a hybrid prescription in which the homogeneous solution is retained only when the inflaton is not thermalized ($n=0$), whereas the quantum stochastic source is restored when Bose--Einstein statistics are assumed for the inflaton fluctuations ($n=n_{\rm BE}$). \texttt{SWIM} and \texttt{DSWIM} employ the quantum stochastic source together with vanishing initial conditions, irrespective of the choice of inflaton statistics. 

These different prescriptions differ primarily in the moderate dissipative regime, while remaining in close agreement in the weak- and strong-dissipation limits, as observed in ref.~\cite{Ballesteros:2023dno} and in the numerical comparisons presented in section~\ref{sec:Results} of this work. Consequently, the moderate-$Q$ discrepancy between \texttt{WI2easy} and \texttt{SWIM}/\texttt{DSWIM} should be interpreted as arising from the different prescriptions adopted for incorporating the quantum contribution to the primordial fluctuations.

To facilitate direct comparisons with \texttt{WI2easy}, we have introduced an optional \texttt{wi2easy} flag in both \texttt{SWIM} and \texttt{DSWIM}. When enabled (\texttt{wi2easy}=1), the codes reproduce the hybrid prescription adopted in \texttt{WI2easy}. When disabled $(\texttt{wi2easy}=0$, the default setting), \texttt{SWIM} and \texttt{DSWIM} employ the prescription used throughout this work, in which the quantum stochastic source is always retained and the inflaton perturbations are initialized with vanishing initial conditions.



\acknowledgments
U.K. acknowledges Suratna Das for numerous insightful discussions and suggestions that directly contributed to this work. U.K. is grateful to Rudnei O. Ramos, Alejandro Perez Rodriguez, and Gabriel Rodrigues for valuable correspondence, helpful discussions, and comments on the deterministic formulation and numerical implementation. U.K. also acknowledges the Axis Bank PhD Program at Ashoka University for the PhD fellowship provided by Axis Bank.

\bibliographystyle{JHEP}
\bibliography{ref.bib}

\end{document}